\documentclass{aa}
\usepackage{graphicx}
\usepackage{txfonts}
\usepackage{color}
\usepackage{xcolor}
\usepackage{siunitx}
\usepackage[hidelinks]{hyperref}
\hypersetup{
    colorlinks,
    linkcolor={red},
    citecolor={blue!80!black},
    urlcolor={blue!80!black}
}
\usepackage{pifont}
\newcommand{\cmark}{\ding{51}}%
\newcommand{\xmark}{\ding{55}}%
\newcommand{\molecfit}{\textit{Molecfit}}

\begin{document}

\title{Telluric correction in the near-infrared:\\
  Standard star or synthetic transmission?}
\titlerunning{ }
\author{S. Ulmer-Moll\inst{\ref{inst1},\ref{inst2}},
  P. Figueira\inst{\ref{inst3}, \ref{inst1}},
  J. J. Neal\inst{\ref{inst1},\ref{inst2}},
  N. C. Santos\inst{\ref{inst1},\ref{inst2}}
  and M. Bonnefoy\inst{\ref{inst4}}}

\institute{
  Instituto\,de\,Astrofísica\,e\,Ciências\,do\,Espaço,\,Universidade\,do\,Porto,\,CAUP,\,Rua\,das\,Estrelas,\,4150-762\,Porto,\,Portugal\label{inst1}
  \and
  Departamento de Física e Astronomia, Faculdade de Ciências, Universidade do Porto, Portugal\label{inst2}
  \and
  European Southern Observatory, Alonso de Cordova 3107, Vitacura, Santiago, Chile\label{inst3}
  \and
  Univ. Grenoble Alpes, CNRS, IPAG, F-38000 Grenoble, France\label{inst4}}

\date{Received ...; accepted ...}

\abstract
% context heading (optional)
    {The  atmospheric absorption of the Earth is an important limiting factor for ground-based spectroscopic observations and the near-infrared and infrared regions are the most affected.
      Several software packages that produce a synthetic atmospheric transmission spectrum
      have been developed to correct for the telluric absorption; these are \molecfit, TelFit, and Transmissions Atmosphériques Personnalisées pour l’AStronomie (TAPAS).}
  % aims heading (mandatory)
    {Our goal is to compare the correction achieved using
      these three telluric correction packages and the division by a telluric standard star. 
      We want to evaluate the best method to correct near-infrared high-resolution spectra
      as well as the limitations of each software package and methodology.}
  % methods heading (mandatory)
    {We applied the telluric correction methods to CRIRES archival data taken in the J and K bands.
      We explored how the achieved correction level varies depending on the atmospheric T-P profile
      used in the modelling, the depth of the atmospheric lines, and the molecules creating the absorption.}
  % results heading (mandatory)
    {We found that the \molecfit\ and TelFit corrections lead to smaller residuals for the water lines.
      The standard star method corrects best the oxygen lines. The
      \molecfit\ package and the standard star method corrections result in global offsets always below 0.5\% for all lines;
      the offset is similar with TelFit and TAPAS for the $H_2O$ lines and around 1\% for the $O_2$ lines.
      All methods and software packages result in a scatter between 3\% and 7\% inside the telluric lines. 
      The use of a tailored atmospheric profile for the observatory leads to a scatter two times smaller,
      and the correction level improves with lower values of precipitable water vapour.}
  % conclusions heading (optional)
    {The synthetic transmission methods lead to an improved correction compared to
      the standard star method for the water lines in the J band with no loss of telescope time,
      but the oxygen lines were better corrected by the standard star method.}

    \keywords{atmospheric effects --
      radiative transfer --
      instrumentation: spectrographs --
      methods: data analysis --
      techniques: spectroscopic
    }

  \maketitle

%
%________________________________________________________________

  \section{Introduction}
  In ground-based observations, the light coming from a celestial object is partly or totally absorbed by the Earth's atmosphere,
  a phenomenon that is strongly wavelength dependent.
  Even if the position of an observatory is carefully chosen to minimize the impact of the atmosphere,
  there is still a need to correct for telluric absorption.
  In spectroscopic studies, the species present in the atmosphere imprint absorption or emission lines on top of the spectra of the target.
  In absorption, the telluric lines create what is called the transmission spectrum of the Earth's atmosphere.
  The volume mixing ratio of the different molecules as a function of height
  (that can be understood in terms of density profile) present in the atmosphere and the atmospheric conditions (pressure, temperature)
  affect the telluric lines in their shape, depth, and position in wavelengths.
  High winds can shift the telluric features,
  \cite{figueira_comparing_2012} showed that a horizontal wind model can account for some of these shifts
  and is in agreement with radiosonde measurements.
  Thus, the transmission spectrum depends strongly on the time and location of the observations.
  Every molecule contributes differently to the final transmission.
  For example, $H_2O$ leads to an absorption over a very wide wavelength range,
  spanning the optical, near-infrared, and infrared.
  This absorption defines the near-infrared bands on which photometry and spectroscopy was performed for many years.
  The water vapour shows hourly to seasonal variations that are challenging to correct \citep{adelman_table_2003,wood_telluric_2003}.
  $O_2$ absorption might be easier to correct because it provides sharp, deep, and well-defined features
  and the $O_2$ volume mixing ratio is more stable in the atmosphere.
  $O_2$ bands and individual lines in the optical have been studied for a long time \citep{wark_absorption_1965,caccin_terrestrial_1985}.
  When observations are done through cirrus clouds, the atmospheric transmission is not impacted.
  Cirrus clouds are thin clouds made of ice crystals and usually found at altitudes higher than 6 km \citep{wylie_four_1994}.
  The ice crystals transmit most of the incoming stellar light and do not introduce narrow water features in the transmission spectrum.

  A correction of telluric absorption is required when the studies aim at high spectral fidelity;
  for example it is an essential step to characterize planetary and exoplanetary atmospheres
  (e.g. \citealt{bailey_correcting_2007, cotton_atmospheric_2014, brogi_carbon_2014}).
  Telluric correction has also been studied in the context of exoplanet search through radial velocity (RV) measurements.

  Earlier on, the telluric lines started to be used as a wavelength calibration
  to measure precise radial velocities \citep{griffin_accurate_1973}.
  The atmosphere can be considered as a gas cell which imposes its absorption features on the target spectrum.
  \cite{smith_precise_1982} and \cite{balthasar_terrestrial_1982} used $O_2$ lines as wavelength calibrators
  and could reach a precision down to 5 ${\rm m}\cdot {\rm s}^{-1}$.
  With UVES data, \cite{snellen_new_2004} used $H_2O$ lines as wavelength reference
  and reached a precision of 5-10 ${\rm m}\cdot {\rm s}^{-1}$.

  \cite{figueira_evaluating_2010} studied the stability of the telluric lines to serve as a wavelength calibration
  and found a stability of 10 ${\rm m}\cdot {\rm s}^{-1}$ over six years and down to 5 ${\rm m}\cdot {\rm s}^{-1}$ over shorter timescales.
  \cite{figueira_radial_2010} then used the telluric lines as wavelength reference to obtain a RV precision
  in the near-infrared of 10 ${\rm m}\cdot {\rm s}^{-1}$ with CRIRES data.
  Using the ammonia gas cell technique as wavelength calibration,
  \cite{bean_crires_2010} were also able to achieve a RV precision of 5 ${\rm m}\cdot {\rm s}^{-1}$
  and down to 3 ${\rm m}\cdot {\rm s}^{-1}$ on short timescales with CRIRES data.

  The achievement of higher RV precision implies more stringent limits on the spectral fidelity.
  This is particularly true for the visible wavelengths, in which a submetre per second is already achieved today \citep[e.g.][]{lovis_exoplanet_2006}.
  In the visible wavelengths, \cite{cunha_impact_2014} looked at the impact of micro telluric lines (with a depth up to 2\%)
  and demonstrated that the RV impact of these telluric lines was in the range of 10 - 20 ${\rm cm}\cdot {\rm s}^{-1}$,
  which is comparable to the RV precision that is expected to be achieved by ESPRESSO
  \citep{pepe_espresso_2014, gonzalezhernandez_espresso_2017}
  and required to detect an Earth-mass planet around a solar-like star.
  In the near-infrared wavelengths, \cite{figueira_radial_2016} presented calculations of the RV information content
  and showed that the telluric lines are limiting the achievable RV precision but without quantifying the impact.
  It was found that the impact on RV precision depended on multiple parameters.
  Some were rooted in the specifications of the spectrograph, like resolution or line sampling.
  Others were related to the level or quality of the correction itself.
  However, the final precision depended on how the telluric and stellar spectra overlapped.
  This means that even for fixed instrumental properties and correction precision,
  the achievable RV depends significantly on the stellar type, systemic RV, and v.sini.
  
  The historical method to correct for the telluric absorption is the standard star method,
  where the spectrum of the target is divided by the spectrum of a fast-rotating hot star (often of type O, B, or A)
  called a telluric standard star \citep{vidal-madjar_deuterium_1986,vacca_method_2003}.
  The high temperature and fast rotation of the standard star lead to a low number of resolved stellar lines,
  so that its spectrum accurately reproduces the atmospheric transmission spectrum.
  The remaining stellar features are broadened by stellar rotation contrary to the telluric lines which remain narrower,
  allowing for an easy identification and second-stage removal.
  The division by the spectrum of the standard star also removes the blaze effect
  introduced by the spectrograph grating, but this effect can be removed during the data reduction process by other means.
  In order to accurately reproduce the atmospheric transmission spectrum,
  the standard star spectrum needs to be taken close in airmass and in time to the spectrum of the target.
  For reference, at ESO the telluric reference stars are only considered suitable for correction
  if and only if taken within 2 h and at an airmass difference of up to 0.2 relative to the target star.
  
  However, there are several fundamental limitations in the telluric correction level
  that can be achieved with the standard star method.
  First, stellar features can still be present in the spectrum of the standard star.
  The time of the observation and the pointing direction between the two observations are not the same,
  which means the light paths probed in the atmosphere are not the same,
  and as such the atmosphere imprint on these two observations is different \citep{lallement_correction_1993,wood_telluric_2003}.
  \cite{bailey_correcting_2007} showed that the telluric correction of the spectrum of a Sun-like star
  with the standard star method can lead to errors up to a few percent for the strong telluric lines,
  and that these errors can be up to 50\% when correcting the observations of $CO_2$ features in the atmosphere of Mars.

  The authors also demonstrated that these errors can be reduced when a synthetic spectrum of the telluric transmission,
  calculated with a radiative transfer model, is used instead of a standard star spectrum.
  \cite{lallement_correction_1993} performed the first telluric correction
  with forward modelling for the $H_2O$ lines near the sodium doublet at \SI{589.5}{\nano\metre}.
  \cite{seifahrt_synthesising_2010} reproduced observed telluric lines with a synthetic transmission spectrum
  to down to a correction level of around $2\%$.
  In the last few years, the telluric correction
  with synthetic transmission spectrum of the Earth's atmosphere has been intensively developed.
  TelFit\footnote{\href{https://pypi.org/project/TelFit/}{pypi.org/project/TelFit}} \citep{gullikson_correcting_2014},
  Transmissions Atmosphériques Personnalisées pour l’AStronomie\footnote{\href{http://cds-espri.ipsl.fr/tapas/}{cds-espri.ipsl.fr/tapas}} (TAPAS; \citealt{bertaux_tapas_2014}),
  and \molecfit \footnote{\href{http://www.eso.org/sci/software/pipelines/skytools/molecfit}{eso.org/sci/software/pipelines/skytools/molecfit}} \citep{smette_molecfit_2015, kausch_molecfit:_2015}
  among others \cite[e.g.][]{rudolf_modelling_2016,villanueva_planetary_2018} are publicly available codes or web interfaces to perform telluric correction.
  
  Other techniques that retrieve the telluric spectra directly from the observations
  (in opposition to forward modelling) have been successful,
  by making clever use of archival data or the sequence of the observations.
  \cite{artigau_telluricline_2014} presented a method based on principal component analysis,
  which uses a large number of standard star observations to build a library of telluric features.
  The telluric features in the spectra of interest are matched with those in the library and removed.
  \cite{astudillo-defru_groundbased_2013} and \cite{wyttenbach_spectrally_2015}
  looked at the variation of the flux with the airmass to identify the telluric lines
  and build a telluric spectrum from the target observations themselves.
  By assuming a linear dependence between airmass and telluric absorption on the affected pixels
  (stable in the rest-frame of the detector), they are able to correct high-resolution spectra
  and detect the elusive calcium and sodium absorption features (respectively) in exoplanetary atmospheres.
  \cite{snellen_groundbased_2008} and \cite{brogi_signature_2012, brogi_carbon_2014} had already used a similar approach
  to detect sodium, carbon monoxide, and water vapour in extrasolar planet atmospheres.

  On the other hand, \molecfit\ has been used to search for water vapour on exoplanet atmospheres \citep{allart_search_2017a}
  and to study the atmospheric conditions above ESO's new astronomical site Cerro Amazones \citep{lakicevic_atmospheric_2016}.
  The quest for Earth-like planets around M dwarf drives the development of infrared high-resolution spectrographs.
  With the arrival of a large number of these spectrographs, such as SPIRou \citep{moutou_spirou_2015},
  CRIRES+ \citep{follert_crires_2014}, NIRPS \citep{wildi_nirps_2017}, and CARMENES \citep{quirrenbach_carmenes_2014} among others,
  the improvement of the ability to perform telluric correction is becoming inevitable.

  The different atmospheric models and methodologies should be evaluated in order to understand what 
  is the most efficient way of correcting the absorption introduced by the Earth's atmosphere.
  To do so, in the context of RV searches,
  we used archival CRIRES data and focussed our experiment on a short wavelength domain under well-characterized environmental parameters.
  In Section \ref{obs}, we introduce the CRIRES data and the reduction process we used.
  In Section \ref{tell_corr}, we present the three correction methods (\molecfit, TelFit, and TAPAS)
  and how we implemented the correction procedure in each of them.
  Section \ref{comparison} presents the correction done with the standard star method.
  This Section also shows the comparison of the telluric corrections and the impact of airmass
  and the atmospheric profile on the correction quality.
  A discussion of the results and the conclusions are presented in Section \ref{conclusion}.

%
%________________________________________________________________

\section{Observations and reduction}
\label{obs}
To test the telluric correction methods we used near-infrared observations from CRIRES,
a high-resolution spectrograph installed at the the Very Large Telescope (VLT) in Paranal.
It covers the infrared region from 1.0 to \SI{5.3}{\micro\metre} with a resolving power up
to 100,000 depending on the slit being used \citep{kaeufl_crires_2004}.
The simultaneous wavelength coverage is however small
from $\lambda / 70$ at \SI{1}{\micro\metre} to $\lambda / 50$ at \SI{5}{\micro\metre}.
We describe below the observations and data reduction.

\subsection{Observations}

\begin{figure}
  \resizebox{\hsize}{!}{\includegraphics{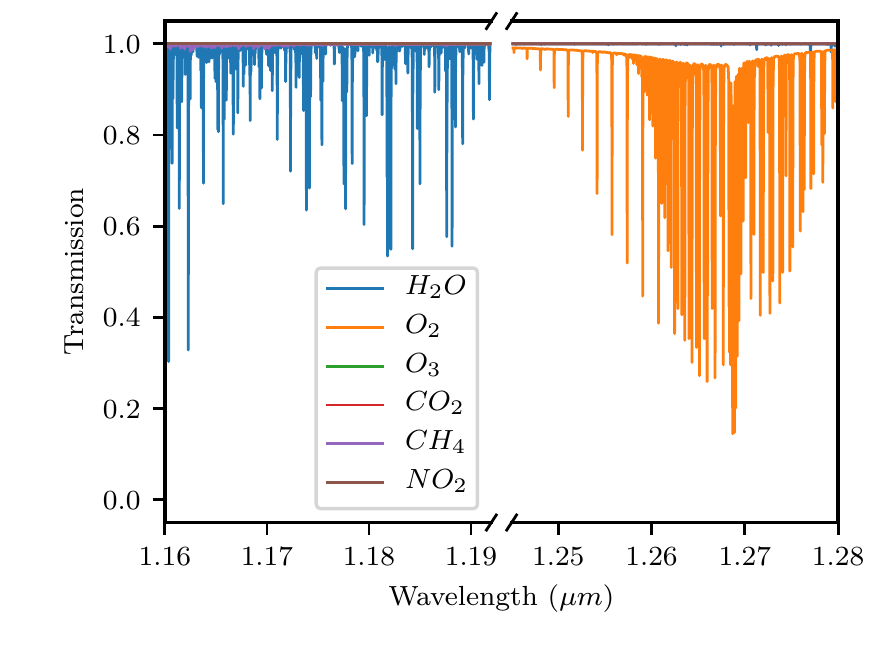}}
  \caption{Synthetic transmission spectrum of our atmosphere calculated with the LBLRTM through the TAPAS interface. The main absorber in the first wavelength range is $H_2O$ (blue line), while in the second range it is $O_2$ (orange line). The other molecules have a transmission very close to 1.}
  \label{fig:transmission}
\end{figure}

We used archival CRIRES spectra of an M7 V - M8 V dwarf star \citep{lowrance_candidate_2000},
HR 7329-B, and a telluric standard of spectral type B9 V \citep{houk_michigan_1978}, HIP100090.
According to the CRIRES User Manual version 84.2\footnote{\href{https://www.eso.org/sci/facilities/paranal/instruments/crires/doc/VLT-MAN-ESO-14500-3486_v84.pdf}{eso.org/sci/facilities/paranal/instruments/crires/doc/VLT-MAN-ESO-14500-3486\_v84.pdf}},
the value of the resolution (R) for other slit widths (sw in arcseconds)
are derived with the following relation: $\frac{R * sw}{0.2} \approx 100,000$.
The 0.4'' slit width was used, which leads to a resolution around 50,000.
The observations were done in two orders.
The first order spans from \SI{1.166}{\micro\metre} to \SI{1.188}{\micro\metre}
and $H_2O$ dominates the telluric absorption in this order.
The second order spans from \SI{1.247}{\micro\metre} to \SI{1.272}{\micro\metre}
with $O_2$ dominating the telluric absorption in this region.
The atmospheric transmission in the wavelength regions we consider is shown in Figure~\ref{fig:transmission}.
The observations are listed in Table \ref{table:obs}.
The spectra were acquired on three consecutive nights from 4 to 6 August 2009 and again from 19 to 21 September 2009 for the two stars,
observing the telluric standard less than 2 h after the target star.
The signal-to-noise ratio (S/N) is calculated with the Beta Sigma noise estimation from \cite{czesla_posteriori_2018}.
We followed their noise estimation on the high-resolution spectrum with several narrow spectral features
because this is also the case for our target star and we chose a jump parameter of 2 and an order of approximation of 5.

\begin{table*}
  \caption{Details of the observations. The airmass column contains the value at the start and end of the observations.
    The seeing value is corrected for the airmass and corresponds to the seeing at the zenith.}
  \label{table:obs}
  \centering
  \begin{tabular}{l c c c c c c c}
    Star                           & Wavelength                & Date and time            & Airmass         & Seeing  & Exp time              & S/N     & OB number\\
                                   & (\SI{}{\micro\meter})     &                          &                 &         & (s)                   &         & \\
  \hline
  \noalign{\smallskip}
  HIP 100090                       & 1.166 - 1.188             & 2009-08-04 03:19         & 1.131 - 1.134   & 0.50     & 180.                 & 360     & (...)   \\
                                   &                           & 2009-08-05 04:19         & 1.118 - 1.118   & 0.62     & 180.                 & 265     & (...)    \\
                                   &                           & 2009-08-06 04:16         & 1.118 - 1.119   & 0.64     & 180.                 & 300     & (...)      \\
  HR 7329-B                        &                           & 2009-08-04 02:23         & 1.165 - 1.152   & 0.56     & 300.                 & 45      & 1     \\
                                   &                           & 2009-08-05 03:22         & 1.153 - 1.175   & 0.67     & 300.                 & 37      & 2      \\
                                   &                           & 2009-08-06 03:23         & 1.154 - 1.180   & 0.75     & 300.                 & 30      & 3       \\
  HIP 1000090                      & 1.247 - 1.272             & 2009-09-19 02:24         & 1.157 - 1.161   & 0.74     & 180.                 & 416     & (...)    \\
                                   &                           & 2009-09-20 01:33         & 1.122 - 1.124   & 0.70     & 180.                 & 434     & (...)     \\
                                   &                           & 2009-09-21 01:09         & 1.117 - 1.118   & 0.75     & 180.                 & 400     & (...)       \\
  HR 7329-B                        &                           & 2009-09-19 01:30         & 1.192 - 1.199   & 0.77     & 300.                 & 45      & 4      \\
                                   &                           & 2009-09-20 00:40         & 1.158 - 1.161   & 0.67     & 300.                 & 55      & 5       \\
                                   &                           & 2009-09-21 00:17         & 1.153 - 1.154   & 0.67     & 300.                 & 52      & 6        \\
  \end{tabular}
\end{table*}

\subsection{Data reduction}

The data was reduced using a custom IRAF\footnote{IRAF is distributed by the National Optical Astronomy Observatories,
  which are operated by the Association of Universities for Research in Astronomy, {Inc.},
  under cooperative agreement with the National Science Foundation.}
pipeline described in detail in \cite{figueira_radial_2010}.
The reduction steps covered are mutual subtraction of eight ABBABBA nodding pairs, linearity correction, and optimal extraction.
The extracted spectra are individually continuum normalized,
and then co-added to create one spectrum per observation with a higher S/N.
An initial wavelength solution is taken from the header keywords provided by the instrument.

%
%________________________________________________________________

\section{Telluric correction}
\label{tell_corr}

TelFit, \molecfit, and TAPAS are three telluric correction methods based on the computation
of the atmospheric transmission with the radiative transfer code LBLRTM \citep{clough_linebyline_1995, clough_atmospheric_2005}.
This forward code solves the equation of radiative transfer by calculating the contribution of each line individually.
The code uses an atmospheric profile and the line parameters as inputs.
The atmospheric profile describes the temperature, pressure and molecule abundances as a function of the altitude.
The line parameters include position, intensity, and broadening coefficients taken from the spectroscopic databases
HITRAN 2008 \citep{rothman_hitran_2009} and HITRAN 2012 \citep{rothman_hitran2012_2013}.

A summary of the capabilities and models employed by each package is shown in Table \ref{table:features}.
Appendix~\ref{appendix:trans} presents the spectra of the telluric standard
and the synthetic transmission spectra obtained with each of the packages.

\begin{table*}
  \caption{Comparison of the modelling parametrization of the telluric correction methods. \molecfit\ and TelFit packages allow for fitting the atmospheric transmission to the target spectrum.}
  \label{table:features}                          % is used to refer this table in the text
  \centering                                      % used for centering table
  \begin{tabular}{l c c c}                        % centered columns (4 columns)
  \hline                                          % inserts single horizontal line
  \noalign{\smallskip}
  Characteristics                                &\molecfit             & TelFit            & TAPAS            \\
  \noalign{\smallskip}\hline\noalign{\smallskip}
  Wavelength coverage (\SI{}{\micro\meter})      & 0.3-30               & 0.3-2.4  & 0.35-2.5          \\
  Radiative transfer code                        & LBLRTM               & LBLRTM            & LBLRTM             \\
  Spectroscopic database                         & HITRAN 2008/12       & HITRAN 2008/12    & HITRAN              \\
  Weather models database                        & GDAS                 & GDAS              & ETHER                \\
  %%\hline\noalign{\smallskip}
  Default atmospheric profile                    & Merged                    & Mid-latitude      & Arletty                 \\
  Other atmospheric profiles                     & Equatorial, Mid-latitude  & GDAS              & Equatorial, Mid-latitude \\
  Automatic download profile                     & \cmark                & \xmark                 & \xmark                         \\
  Fitting atmospheric transmission               & \cmark                & \cmark        & \xmark                          \\
  \noalign{\smallskip}\hline\noalign{\smallskip}
  Atmospheric transmission options:              &                       &                   &                          \\ 
  Molecules: $H_2O,CO_2,O_3,N_2O,CH_4,O_2$        & \cmark            & \cmark        & \cmark                \\
  $CO, NO,SO_2,NO_2, NH_3,HNO_3 $                 & \cmark +11 others & \cmark        & \xmark                         \\
  Transmission spectrum of individual molecule   & \xmark                     & \xmark                 & \cmark              \\
  Resolution                                     & \xmark                     & \cmark        &  \cmark              \\
  Gaussian line shape                             & \cmark            & \cmark        & \cmark                \\
  Lorentzian line shape                           & \cmark            & \xmark                 & \xmark               \\
  Voigt line shape                                & \cmark            & \xmark                 & \xmark                \\
  Singular value decomposition                   & \xmark                     & \cmark        & \xmark                 \\
  \noalign{\smallskip}\hline\noalign{\smallskip}
  Fitting parameters:                            &                       &                   &                         \\
  Molecule                                       & \cmark            & \cmark        & (...)                         \\
  Resolution/line shape                           & \cmark            & \cmark        & (...)                        \\
  Wavelength correction                          & \cmark            & \cmark        &  (...)                     \\
  Continuum correction                           & \cmark            & \cmark        &  (...)                      \\
  Telescope background emission                  & \cmark            & \xmark                 &  (...)                       \\
  \hline\noalign{\smallskip}
  \end{tabular}
\end{table*}

\subsection{TelFit}
The Python package TelFit \citep{gullikson_correcting_2014}
models the atmospheric transmission and fits it to a spectrum with the Levenberg-Marquardt algorithm.
TelFit can adjust the wavelength solution of the data to the telluric lines or vice versa with a polynomial function.
The order of the polynomial can be set by the user, but the values of the coefficients have to be directly set in the code.
TelFit offers a parametric model of the line shape through a Gaussian function and a non-parametric model with the singular value decomposition mode.
For each observation, TelFit uses, by default, an average atmospheric profile for mid-latitudes.
The website of the Global Data Assimilation System (GDAS)\footnote{\href{https://ready.arl.noaa.gov/READYamet}{ready.arl.noaa.gov/READYamet}}
enables the  use of a custom atmospheric profile, adapted to the location and time of the observations.
The GDAS profiles are available for observations taken since 2004 and are updated every three hours.

We used the atmospheric profile that was closest to our observations, with a time difference inferior to 2 h, to first fit the spectrum of the standard star.
We experimented with the parametric and non-parametric options for the line shape and attained the same level of correction.
Since the other models studied in this work use parametric correction,
and to provide a more meaningful comparison, we selected the parametric correction.
To perform the fit of the standard star, we left the parameters of the wavelength solution, the continuum level,
line shape, resolution, and the mixing ratio of the molecule
relevant for the wavelength range considered
($H_2O$ or $O_2$) as free parameters.

TelFit does not allow a different parametrization of spectral orders separated by wavelength gaps.
Therefore, each of the three CRIRES detectors is treated independently because the parameters of the wavelength solution differ from one detector to the other.
As such, for each fit we obtain a different value for the following parameters: molecule abundances, resolution, temperature, and pressure profiles.
We homogenize these parameters by taking their mean values weighted by the inverse of the $\chi^2$ calculated for each detector.
Then, we use the homogenized values of the parameters to compute a final atmospheric transmission for each detector.
We divide the stellar spectrum by the atmospheric transmission to correct for the telluric lines.
After the correction of the standard star, we use the values of the wavelength solution as the starting values
for the fit of the target star and repeat the previous procedure to fit the target's spectrum.

\subsection{\molecfit}

\molecfit\ was developed to correct the telluric lines in the visible
and the infrared \citep{smette_molecfit_2015, kausch_molecfit:_2015}.
It also uses LBLRTM and retrieves automatically the atmospheric profile
at the time and place of the observations
from an ESO repository updated weekly.
However, this repository only contains atmospheric profiles
for a limited number of observatories and the user should ask to add a new location.

Similar to TelFit, \molecfit\ fits a model of the atmospheric transmission
to the observed spectrum by adjusting the wavelength solution, continuum level,
molecule abundances, and line shape.
\molecfit\ can also adjust the telescope emissivity and considers a larger number of molecules.
The telluric lines can be modelled by several functions: Gaussian, Lorentzian, Voigt, or boxcar.
Additionally, the so-called expert mode of the package allows for fitting the wavelength and continuum solution
independently for each of the CRIRES detectors.
\molecfit\ receives as input a parameter file that describes the parameters to fit
and the starting values for each of these.
A practical advantage of the expert mode is that it automatically writes a new parameter file
with the results from the fit that can be reused as input for a new fit, making it easier to perform iterative fits.

To perform the telluric correction with \molecfit,
we used a procedure like that implemented for TelFit.
We corrected the standard star spectrum beforehand and then used the wavelength solution in
the resulting parameter file as the starting value for the fit of the target.
The target spectrum is then fitted twice:
First we let free only the parameters of the wavelength solution;
in a second step, we leave the molecule abundances as free parameters
but we exclude regions of the spectrum that contain stellar lines.
Since with expert mode the three detectors are fitted at the same time,
we do not need to homogenize any parameters and perform a second fit using these.
After the second fit, \molecfit\ divides the spectrum of the target
by the computed atmospheric transmission to correct for the telluric lines.

\subsection{TAPAS}
The TAPAS web interface is described in \cite{bertaux_tapas_2014}.
It also computes the atmospheric transmission with the LBLRTM.
However, the atmospheric profile is derived from the
European Centre for Medium-Range Weather Forecasts (ECMW)\footnote{\href{https://www.ecmwf.int/}{ecmwf.int}},
a weather database updated every 6 h.
The result of a TAPAS query is the atmospheric transmission spectrum
at the time and location of the observations, at a chosen resolution.

There is no fitting procedure included in TAPAS, and thus we had to implement one
in order to obtain a telluric correction comparable to TelFit and \molecfit.
Since TAPAS allows us to download the contribution of each molecule independently,
we chose to scale only the abundance of the main absorber in each order, either $O_2$ or $H_2O$, in order to match the observed spectra.
We downloaded two transmission spectra for each observation:
one with only the main absorber, called $T_m$,
and another transmission spectra with the absorption by the other molecules
($O_3$, $CO_2$, $CH_4$, $N_2O$ and $H_2O$ or $O_2$ depending on the order)
and the Rayleigh scattering, called $T_r$.
The impact of the Rayleigh scattering on our observations at \SI{1.1}{\micro\metre} is less then 1\% of the total transmittance (0.995).

The transmission spectrum can be scaled by scaling the optical depths for optically thin lines
as done in \citet{bertaux_tapas_2014} and \citet{ baker_monitoring_2017}.
The total TAPAS transmittance is then given by \( T_{tapas} = (T_m)^{\alpha}T_r \), with $\alpha$ the scaling factor.

In order to correct for the telluric lines,
we performed a series of simple fits of the TAPAS transmission spectrum $T_{tapas}$ to the CRIRES target spectrum $F_{star}$.
The fits are carried out with the Scipy function \verb|curve_fit| using the Levenberg-Marquardt algorithm.
The procedure is similar to what is done in TelFit and \molecfit.
Since it is well known that the wavelength-calibration solution of high-resolution spectrographs
changes from detector to detector, we chose to fit each of these independently.

We performed a wavelength calibration on both the standard star and the target spectra
using the TAPAS transmission spectrum.
The wavelength solution is fitted with a third degree polynomial
$a + b\lambda + c\lambda^2 + d\lambda^3$.
The starting values for $b$, $c$ and $d$ in the fit of the standard star are set to 0.
The offset $a$ is set to 0.05 to facilitate convergence.
The starting values for the target star are the results of the standard star fit.

Then, we fit the continuum of the TAPAS transmission with a second degree polynomial
and we scaled the transmission spectrum of the main absorber as detailed above.
Since we fit the three detectors independently,
we obtained three values for the scaling factor $\alpha$ for one observation.
To homogenize the values of the scaling, we calculated their weighted mean
using as weights the inverse $\chi^2$ of each detector, as was done for the TelFit correction.
Once the wavelength solution, continuum, and scaling have been fitted,
we divided the science spectrum by the fitted TAPAS transmission to perform the telluric correction.

\begin{figure}
  \resizebox{\hsize}{!}{\includegraphics{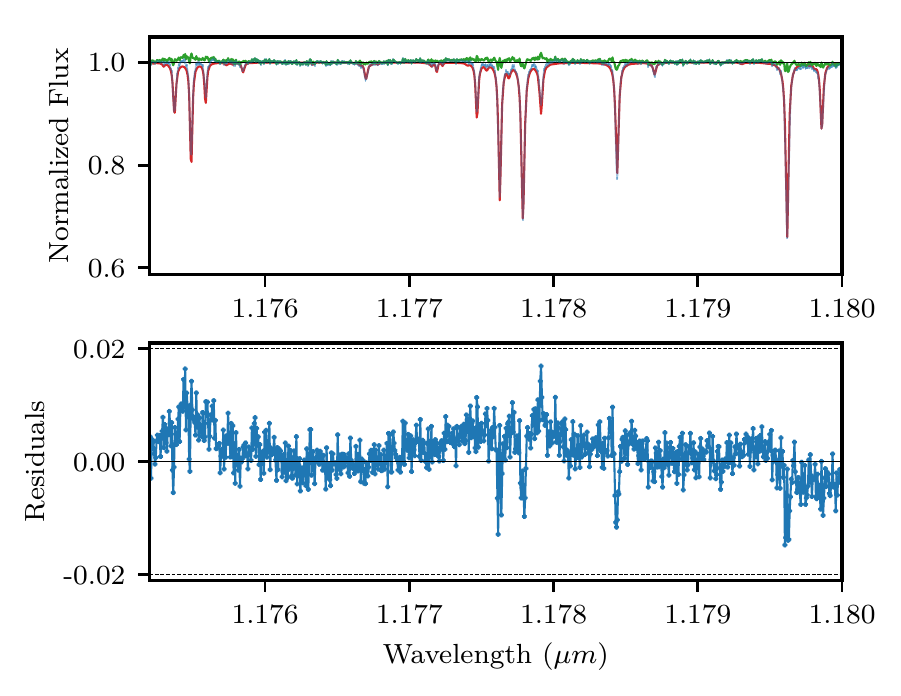}}
  \caption{Standard star spectrum corrected with \molecfit. Top: Input spectrum is indicated in blue, \molecfit\ atmospheric transmission is indicated in red, corrected spectrum is indicated in green. Bottom: Residuals obtained after the subtraction of the atmospheric transmission to the input spectrum.}
  \label{fig:data_exple_std}
\end{figure}

\begin{figure*}
  \centering
  \includegraphics[width=17cm]{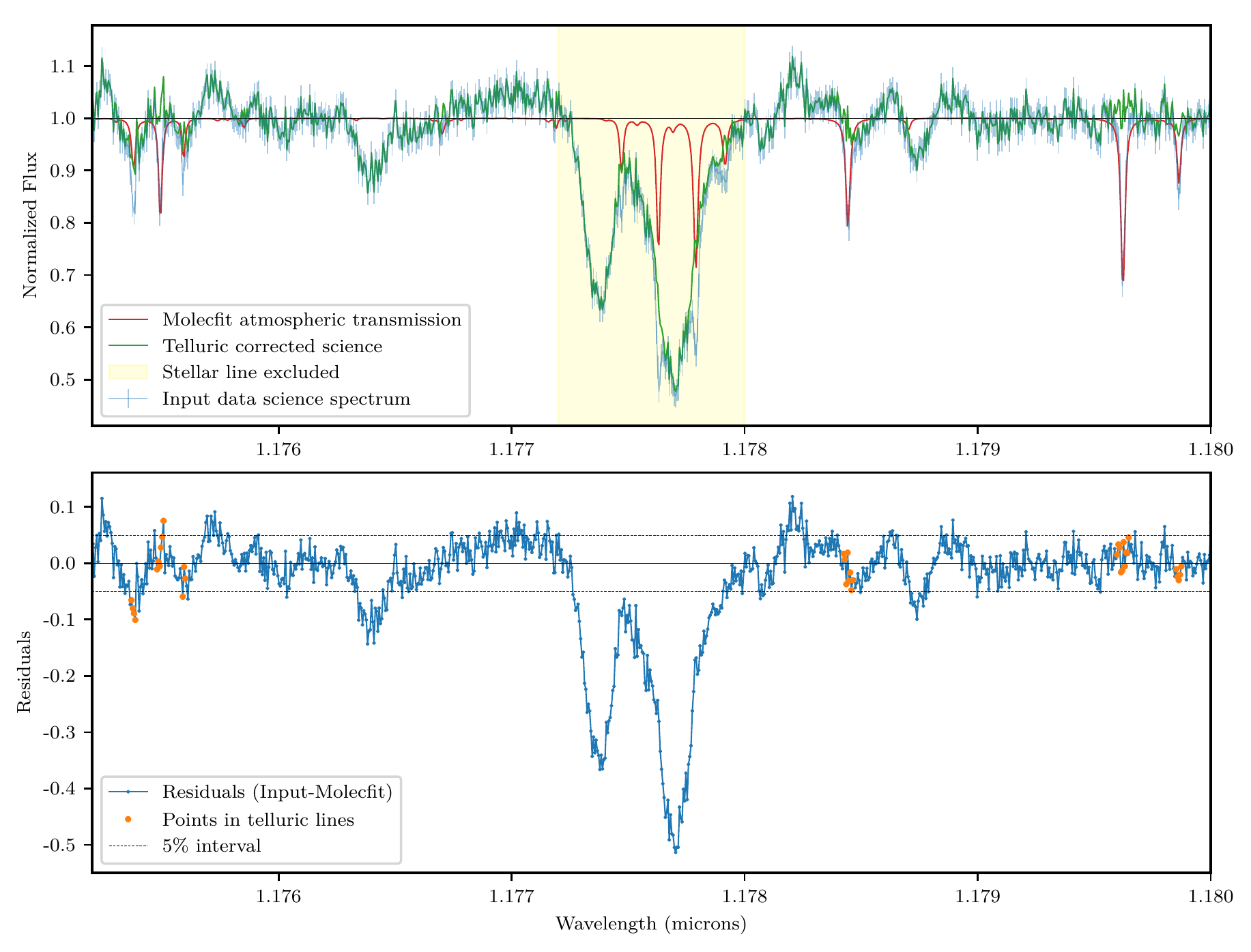}
  \caption{Example of a telluric-corrected spectrum with \molecfit. Top: Input spectrum of the target star is indicated in blue, \molecfit\ atmospheric transmission is shown in red, corrected spectrum is shown in green. Excluded stellar lines are shown in light yellow. Bottom: Residuals between the input spectrum and atmospheric transmission. The dashed lines represent the upper and lower 5\% limits. The points that belong to telluric lines and are not blended with stellar lines are identified in orange.}
  \label{fig:data_exple}
\end{figure*}

%
%
%________________________________________________________________

\section{Comparison}
\label{comparison}

We corrected the CRIRES data with TelFit, \molecfit, and TAPAS following the procedures explained in Section~\ref{tell_corr}.
Since we have standard star observations, we also performed the telluric correction using the standard star method.
We shifted the standard star in RV to match the target star using the \texttt{DopplerShift} function
from the PyAstronomy package\footnote{\href{https://pypi.org/project/PyAstronomy/}{pypi.org/project/PyAstronomy}}.
We chose the RV shift that minimizes the residuals between the two spectra.
Then, we divided the target by the standard star to obtain the telluric-corrected spectrum.

\subsection{Examples of correction}

Two examples of the level of telluric correction obtained with our dataset are presented in this section.
The telluric correction for the whole wavelength range with the four methods is presented in Appendix~\ref{appendix:trans2}.
In Figure~\ref{fig:data_exple_std}, we present an example of the telluric correction with \molecfit, on the third detector, for the standard star spectrum.
The wavelength range is dominated by $H_2O$ absorption and the correction level is high with residuals below 2\%.
Figure~\ref{fig:data_exple} shows the telluric correction of the target spectrum on the same wavelength range.
The telluric correction was also performed with \molecfit. The input spectrum is plotted in blue with the error bars,
the fitted \molecfit\ transmission is in red and shows that all the telluric lines are well matched.
The telluric lines are corrected at the noise level of the spectrum.
In the bottom part, we show the residuals of the difference between the input spectrum and the atmospheric transmission.
We identified strong K I stellar lines with the NIST database\footnote{\href{https://physics.nist.gov/PhysRefData/ASD/lines_form.html}{physics.nist.gov/PhysRefData/ASD/lines\_form.html}}
and we excluded these regions in order to calculate the mean and standard deviation of the residuals
presented in Figures~\ref{fig:all_48} and \ref{fig:all_45}.
In Figures~\ref{fig:tell_48} and \ref{fig:tell_45} we selected only points inside the telluric lines,
thus the broad stellar variations are excluded from the standard deviation measurements.

\subsection{$H_2O$ absorption}
\begin{figure}
  \resizebox{\hsize}{!}{\includegraphics{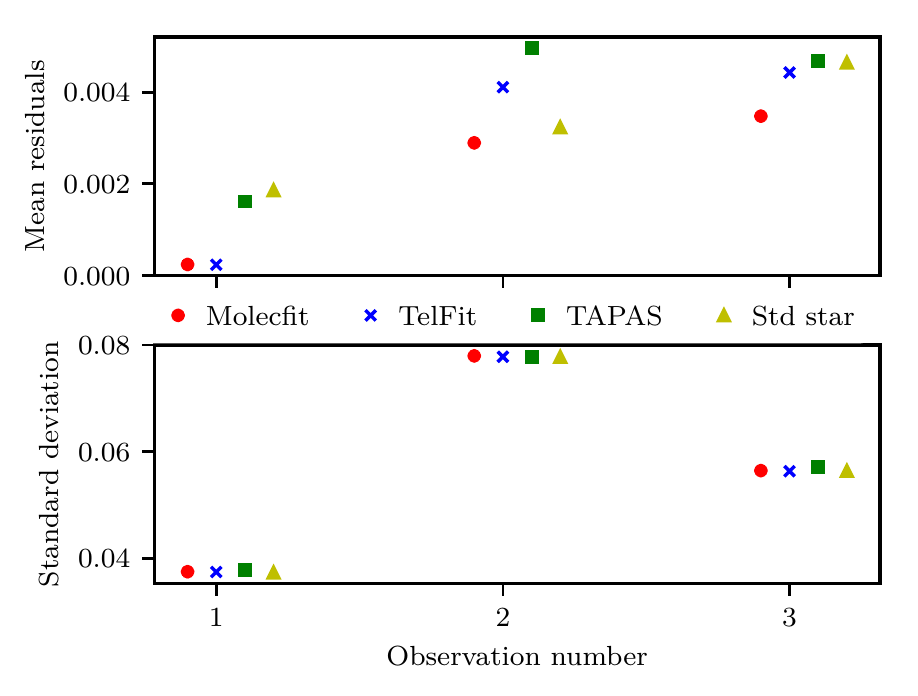}}
  \caption{Mean and standard deviation of the residuals of all points except the stellar lines for the four telluric correction methods \molecfit\ (red circle), TelFit (blue cross), and TAPAS (green square), and the standard star method (yellow triangles) in the wavelength range dominated by $H_2O$ absorption.}
  \label{fig:all_48}
\end{figure}

We compare the telluric correction obtained with TelFit, \molecfit, TAPAS, and the standard star method
in the wavelength range 1.166 to $\SI{1.188}{\micro\meter}$ where $H_2O$ is the main absorber.
Figure~\ref{fig:all_48} presents the mean and standard deviation of the residuals for each of the three observations.
Ideally, the offset of the residuals, represented by the mean, should be zero.
This is the case for TelFit and \molecfit\ for the first observation, which has an offset very close to zero.
For the two other observations, TelFit and \molecfit\ have an offset around $0.3\%$ and $0.4\%$, respectively. The
TAPAS package and the standard star method have a higher offset for the first observation of
close to 0.2\% and a similar offset (around 0.3-0.4\%) for observations 2 and 3.
A positive mean can indicate that the telluric lines might be under-corrected,
meaning the modelled line is not as deep as the line in the input spectrum.
The standard deviation of the residuals should also tend to zero,
however we are limited here by a large scatter already present in the spectra.
The standard deviation of the residuals seems to be only affected by the observation and not by which method is used for the correction.

\begin{figure}
  \resizebox{\hsize}{!}{\includegraphics{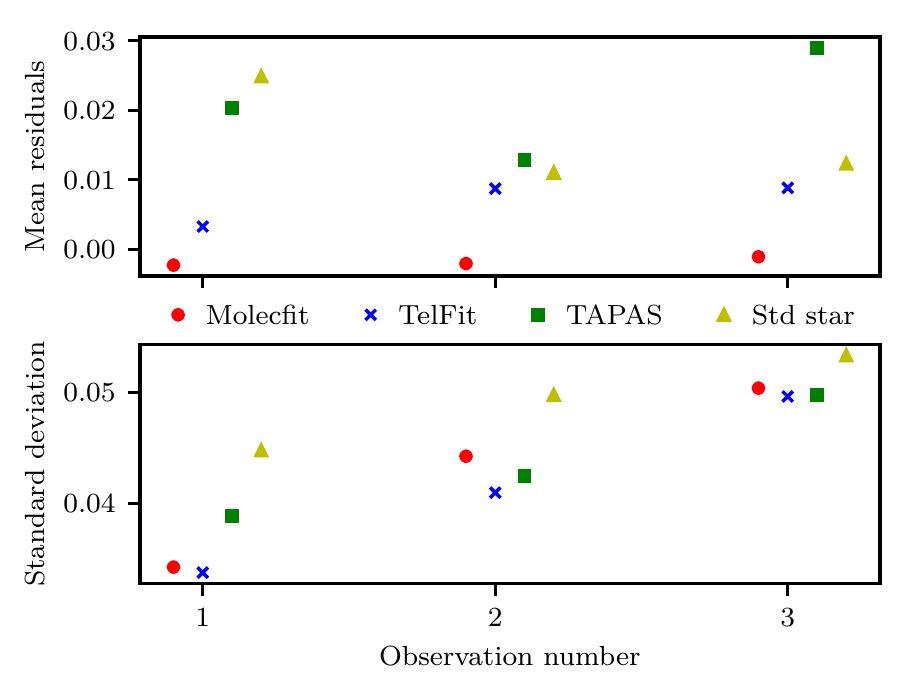}}
  \caption{Mean and standard deviation of the residuals inside the telluric lines for the four telluric correction methods in the wavelength range dominated by $H_2O$ absorption. The colour code is the same as Figure~\ref{fig:all_48}.}
  \label{fig:tell_48}
\end{figure}

Figure~\ref{fig:tell_48} is similar to the previous figure except with the points inside the telluric lines.
\molecfit\ and TelFit have systematically lower scatter and offset than the standard star method. The
\molecfit\  offset is the closest to zero. Also, TAPAS has a lower scatter than the standard star method,
but its offset is higher for observations 2 and 3.
The correction with standard star method could be improved by scaling its spectrum in order to adjust the $H_2O$ absorption.
In this wavelength domain, dominated by water absorption,
we conclude that \molecfit, TelFit, and TAPAS give a better telluric correction than the standard star method.

\subsection{$O_2$ absorption}
\begin{figure}
  \resizebox{\hsize}{!}{\includegraphics{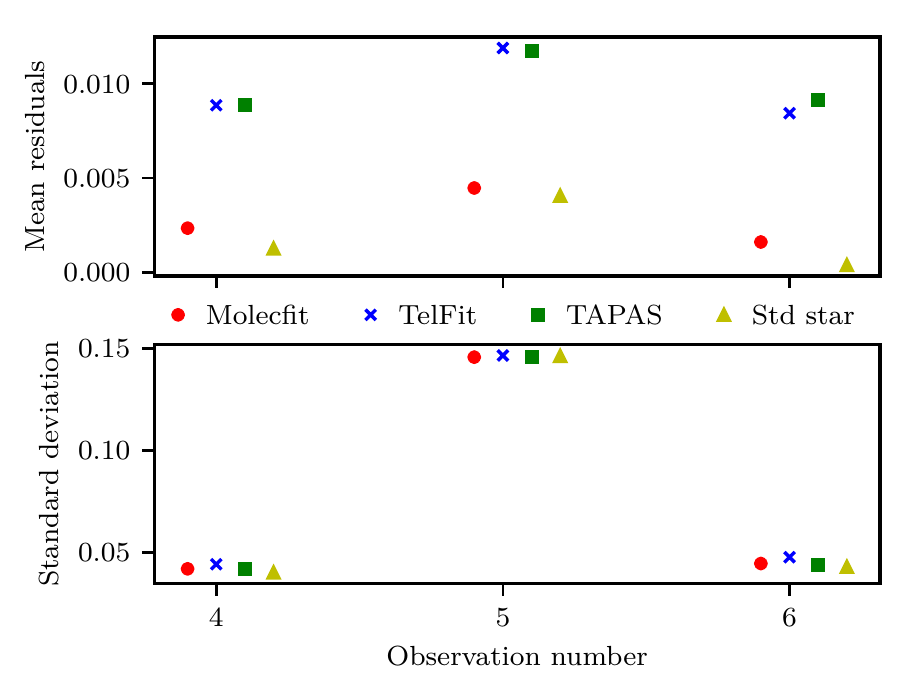}}
  \caption{Mean and standard deviation of the residuals of all points except the stellar lines for the four telluric correction methods in the wavelength range dominated by $O_2$ absorption. The colour code is the same as Figure~\ref{fig:all_48}.}
  \label{fig:all_45}
\end{figure}

\begin{figure}
  \resizebox{\hsize}{!}{\includegraphics{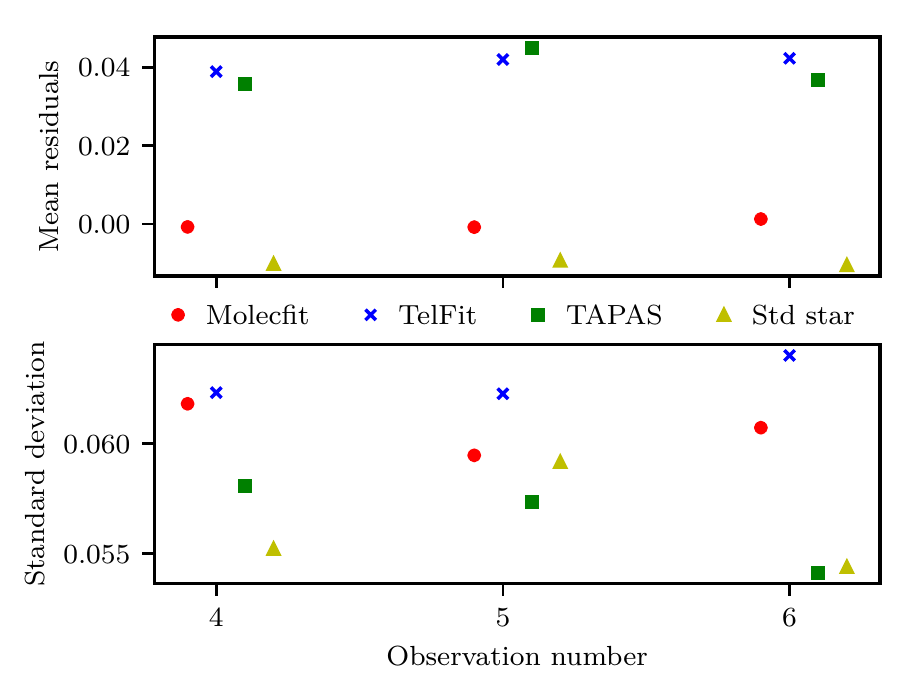}}
  \caption{Mean and standard deviation of the residuals inside the telluric lines for the four telluric correction methods in the wavelength range $O_2$ absorption. The colour code is the same as Figure~\ref{fig:all_48}.}
  \label{fig:tell_45}
\end{figure}

We compare the telluric correction obtained with the four methods
in the wavelength range 1.247 to $\SI{1.272}{\micro\meter}$ where $O_2$ is the main absorber.
Figures~\ref{fig:all_45} and \ref{fig:tell_45} present the offset and scatter of the residuals
for all the points (excluding the K I stellar lines) and for the points inside the telluric lines, respectively.
\molecfit\ and the standard star method clearly give smaller offsets, always below $0.5\%$, for all observations (Fig~\ref{fig:all_45}).
The scatter of the residuals is equivalent for all methods.
For the residuals inside the telluric lines (Fig~\ref{fig:tell_45}),
the smallest offsets are seen with \molecfit\ and the standard star method.
Even though the mean of the \molecfit\ residuals inside telluric lines is the closest to zero,
\molecfit\ presents a higher scatter inside the telluric lines compared to the standard star method and TAPAS.

The correction of $O_2$ telluric lines is clearly more challenging for TAPAS, \molecfit, and TelFit.
These last two packages scale the $O_2$ content in the atmospheric profile, contrary to TAPAS,
where we only scale the $O_2$ transmission a posteriori and to the standard method, for which there is no scaling.
The $O_2$ absorption at $\SI{1.2}{\micro\meter}$ is composed of several deep and wide features at this resolution,
creating a wide absorption band.
This band was partly removed by the continuum normalization completed prior to the telluric correction.
This could explain why the telluric correction with TAPAS results in offset residuals.
For TelFit this explanation is excluded since we did not compute the continuum contribution of the lines with LBLRTM.

\subsection{Shape of the $O_2$ lines}

\begin{figure}
  \resizebox{\hsize}{!}{\includegraphics{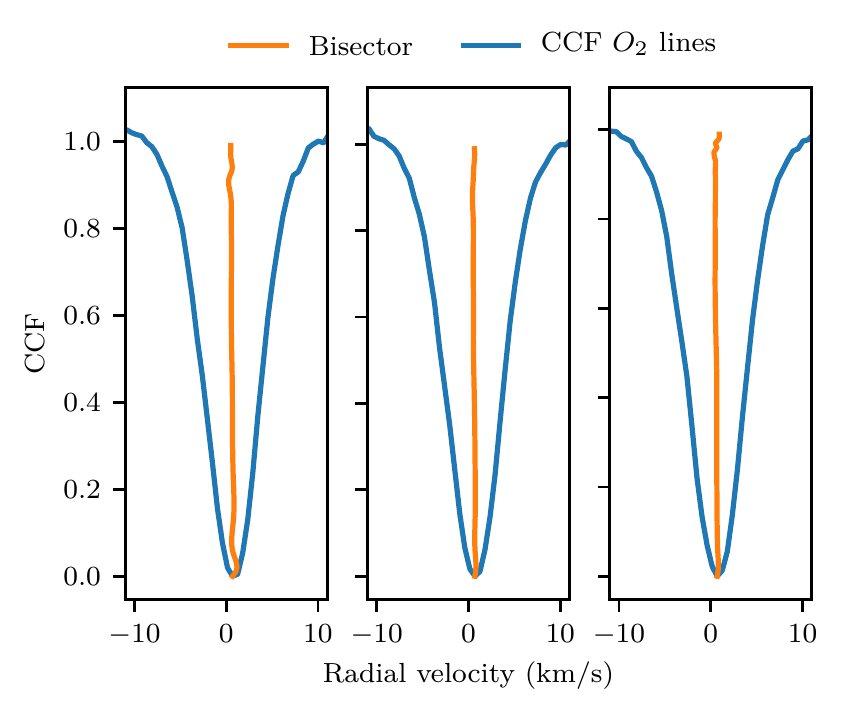}}
  \caption{Cross-correlation function (blue line) and its bisector (orange line) of the $O_2$ lines from the standard star spectra for observations 4, 5, and 6.}
  \label{fig:bisector}
\end{figure}

The line shape is a fitted parameter for the synthetic transmission method.
\molecfit\ and TelFit have several line profiles available, while TAPAS only has a Gaussian profile.
In order to check if the parametrization we used was adequate, we studied the $O_2$ lines imprinted in the standard star spectrum.
To do so, we calculated the cross-correlation function (CCF) between the standard star spectrum and a mask of $O_2$ lines.
The mask contains 22 lines with an average depth of 0.61.
The CCF created by this process has not only a higher S/N, but a much higher sampling than the individual lines present in the original spectra.
As such, this CCF allows a finer evaluation of the average profile of telluric lines.
To create the mask, we selected the positions of the $O_2$ lines from the HITRAN website\footnote{\href{http://hitran.iao.ru/molecule/simlaunch}{hitran.iao.ru/molecule/simlaunch}}.
We calculated the width of the mask in RV  as $\Delta v = \frac{c}{2R} $,
where c is the speed of light, R is the resolution, and the factor of 2 comes from the CRIRES sampling.
Then, we measured the bisector line of the CCF.
The results are plotted in Figure~\ref{fig:bisector}.
We calculated the bisector inverse slope for each observation and we found similar values presented in Table~\ref{table:bis}.
The bisectors calculated for each observation are very similar, which means the variations of the line shape between observations is small.
Besides, the bisector does not show strong asymmetries, which confirms
that the symmetric functions used by \molecfit\ and TelFit to model the lines are adequate.

\begin{table}
  \caption{Bisector inverse slope of the CCF of $O_2$ lines for the three observations.} 
  \label{table:bis}
  \centering
  \begin{tabular}{l c c c}
    Technique                          & OB 1    & OB 2    & OB3     \\
    \hline\\[-0.5em]
    BIS (${\rm km}\cdot {\rm s}^{-1}$)  & -0.452  & -0.480  & -0.488   \\
  \end{tabular}
\end{table}

\subsection{Sensitivity to airmass and relative humidity}

The atmospheric and observing conditions, such as airmass and relative humidity, affect the transmission spectrum of the atmosphere.
We study how the correction for the atmospheric transmission spectrum is influenced by the airmass and relative humidity values.
First, we use the CRIRES data presented in Section~\ref{obs}, which spans a small range of airmasses and humidities.
We notice a positive correlation between the airmass and the standard deviation
of the residuals for all telluric correction methods.
For the relative humidity, the smallest residuals are found at the highest humidity value.
Thus, the slight change in humidity does not seem to affect our telluric correction.

To investigate a wider range of these two parameters, we use a second CRIRES dataset.
This dataset is composed of seven stars with stellar types ranging from F9 V to K3 V.
The 16 spectra are taken in the K band, around \SI{2.15}{\micro\metre}.
The observations were  carried out between April and August 2012, and cover airmasses from 1.0 to 1.6 and relative humidities from 3\% to 33\%.
We reduced the spectra with the same pipeline used for the first dataset and explained in Section~\ref{obs}.
Then, we performed the telluric correction of each spectrum with \molecfit, TelFit, and TAPAS.
The telluric corrections were done as explained in Section~\ref{tell_corr}, but without using a first approximation for the wavelength solution.
We made sure to measure the correction level of telluric lines that were not blended with stellar lines.

\begin{figure}
  \resizebox{\hsize}{!}{\includegraphics{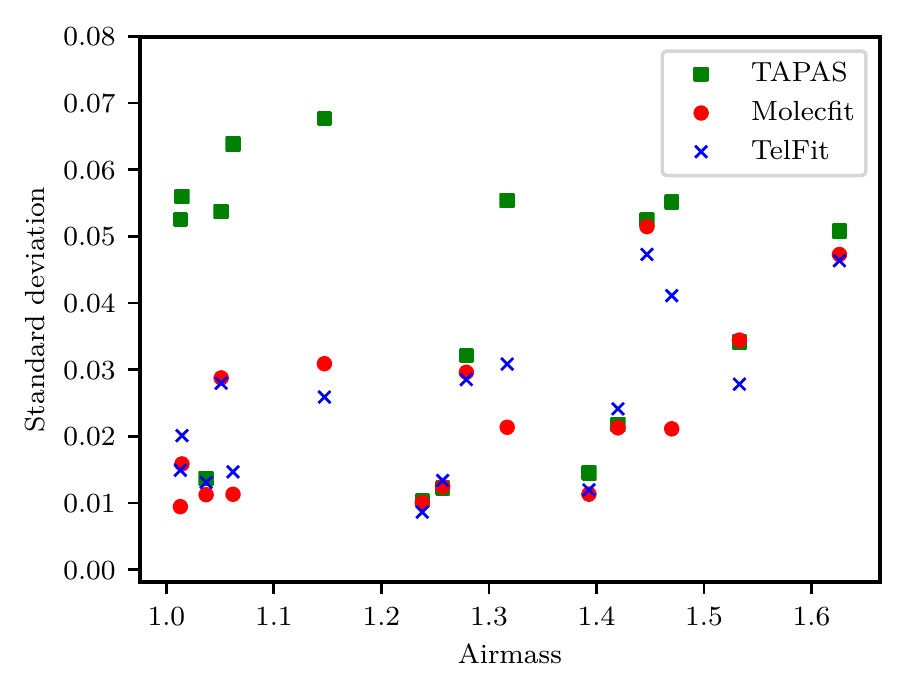}}
  \caption{Scatter of the residuals for the three telluric correction methods \molecfit\ (red circle), TelFit (blue cross), and TAPAS (green square) as a function of airmass.}
  \label{fig:airm_all}
\end{figure}

\begin{figure}
  \resizebox{\hsize}{!}{\includegraphics{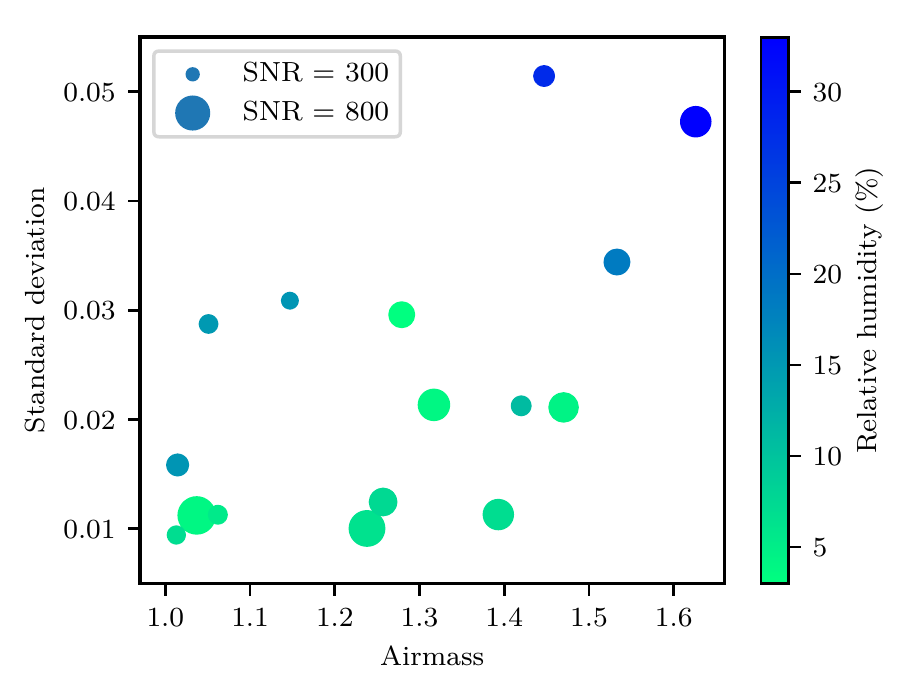}}
  \caption{Scatter of the residuals for the telluric correction with \molecfit\ as functions of airmass (abscissa), relative humidity (colour bar), and S/N (point size).}
  \label{fig:snr_molecfit}
\end{figure}

We show the standard deviation of the residuals, which represents the scatter of the residuals,
obtained with the three methods in Figure~\ref{fig:airm_all}.
Overall, the scatter of the residuals obtained with TAPAS is higher than with \molecfit\ and TelFit,
which confirms the results obtained from the first dataset.
For airmasses below 1.3, the scatter is close to 2\% for \molecfit\ and TelFit while the scatter is around 4\% for TAPAS.
We notice a positive trend of the scatter with airmass for the \molecfit\ and TelFit points but not for the TAPAS points.
In Figure~\ref{fig:snr_molecfit}, we show the residuals of \molecfit\ as functions of airmass, relative humidity, and S/N.
The scatter values obtained with \molecfit\ are similar to those obtained with TelFit.
The relative humidity is traced with the colour bar, the vertical gradient from green to blue indicates
that the scatter increases with higher values of relative humidity.
A positive trend is also noticeable for airmasses higher than 1.4, hence the increasing scatter
can be attributed to both parameters. At low airmass, below 1.2,
the spectra with higher scatter might be explained by their lower S/N and
their relative humidity around 15\%. With this larger sample of airmass and relative humidity,
we find that both of these parameters have an impact on the correction level which can be obtained.
For \molecfit\ and TelFit telluric corrections, higher values of airmass and relative humidity correlate with higher scatter of the residuals.

\subsection{Effect of the atmospheric profile}

We investigate the impact of the input atmospheric profile on the telluric correction.
An atmospheric profile describes how the pressure and temperature vary with altitude,
and the volume mixing ratio of several atmospheric molecules as a function of altitude.
First, we describe three MIPAS atmospheric profiles available as inputs for the telluric correction software packages \molecfit\ and TelFit.
Then we compare the telluric correction obtained by the two profiles available
for \molecfit\ because this package offers the most tailored atmospheric profile for the Paranal Observatory.

\subsubsection{Different atmospheric profiles}
The MIPAS profiles from 2001, created by John Remedios, are available for mid-latitude (\ang{45}), for polar latitude (\ang{75}), and equatorial latitude.
These profiles, available on-line\footnote{\href{http://eodg.atm.ox.ac.uk/RFM/atm/}{eodg.atm.ox.ac.uk/RFM/atm}}, include 30 molecules and 6 others can be included.

The default profile for \molecfit\ is the equatorial daytime profile.
However, \molecfit\ allows the user to use a tailored profile for the Paranal and La Silla observatories
by merging the MIPAS equatorial profile with data from the ESO Meteo Monitor (EMM) and the GDAS profile.
The EMM measures the relative humidity, air temperature, pressure, and dew point temperature close to the telescopes,
and the GDAS profile describes the atmosphere from 0 to 26 km and is updated every 3 h.
For TelFit the default profile is the mid-latitudes at night time profile.
TelFit also facilitates the use of a tailored profile, which is the GDAS profile.

\begin{figure*}
  \centering
  \includegraphics[width=17cm]{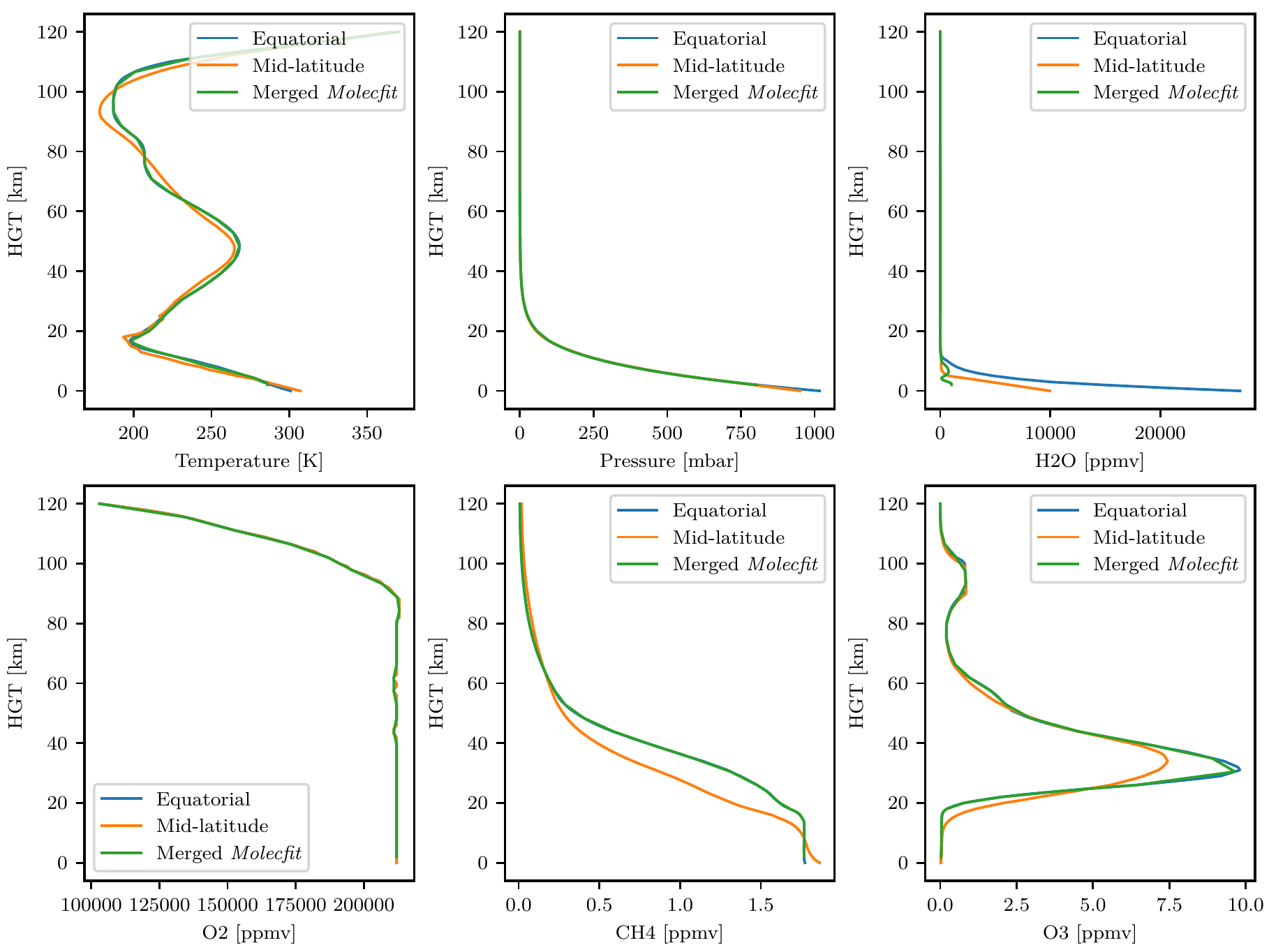}
  \caption{Atmospheric profiles: MIPAS Equatorial in blue, MIPAS mid-latitude daytime in orange, and \molecfit\ best-fit merged profile in green. The merged profile is composed of the interpolation of the GDAS profiles from 05.08.2008 between 3 h and 6 h, and on-site measurements taken in the header of the spectrum. The relative humidity taken at the telescope is equal to 4\%, the pressure to 745 mbar, and the temperature to 286 K.}
  \label{fig:atm_equ_mid}
\end{figure*}

We plot in Figure~\ref{fig:atm_equ_mid} the equatorial daytime profile, mid-latitude profile,
and one of the tailored profiles by \molecfit, called a merged profile.
The $H_2O$ volume mixing ratio of the merged profile is much lower than the two other profiles up to a certain height (16 km),
which reflects the dry condition of the Paranal Observatory.
Using the \molecfit\  merged profile is in this case more adequate to fit the transmission spectrum to the CRIRES data.
The merged profile is identical to the equatorial profile above the altitude of 26 km
because the GDAS profile does not provide information at an altitude higher than 26 km.
The difference between the merged and equatorial are smaller than $10^{-3}$ at an altitude higher than 26 km for all plots.
Below an altitude of 26km, the difference between the merged and equatorial profile are larger.
The median of the difference is equal to 1.9K for the temperature, 1.2 mbar for the pressure,
and 730 ppmv for $H_2O$. For $O_2$, $CH_4$, and $O_3$ the median of the difference is equal to 0.
The $O_2$ volume mixing ratio is similar in the three profiles: equatorial, mid-latitudes, and merged profile.
The volume mixing ratio of $O_3$ and $CH_4$ in the mid-latitude and equatorial profiles differ
but we cannot quantify the differences since $O_3$ and $CH_4$ do not affect the wavelength range of our observations.

Moreover, the concentration of greenhouse gases such as $CH_4$ and $CO_2$ is increasing every year,
as shown on the Earth System Research Laboratory website\footnote{Pieter Tans, NOAA/ESRL, \href{https://www.esrl.noaa.gov/gmd/ccgg/trends/}{esrl.noaa.gov/gmd/ccgg/trends}}
with measurements done at Mauna Loa, Hawaii.
The concentration of $CH_4$ in 2001 (MIPAS profiles) was 1.771 ${\rm ppmv}$ and was 1.842 ${\rm ppmv}$ in 2016,
while the concentration of $CO_2$ was equal to 371.1 ${\rm ppmv}$ in 2001 and increased to  404.21 ${\rm ppmv}$ in 2016.
For that matter the \molecfit\ User Manual\footnote{\href{ftp://ftp.eso.org/pub/dfs/pipelines/skytools/molecfit/VLT-MAN-ESO-19550-5772_Molecfit_User_Manual.pdf}{eso.org/pub/dfs/pipelines/skytools/molecfit/VLT-MAN-ESO-19550-5772\_Molecfit\_User\_Manual.pdf}}
recommends increasing the value of $CO_2$ by $6\%$. 

\subsubsection{Correction with the equatorial and  merged profiles}

\begin{figure}
  \resizebox{\hsize}{!}{\includegraphics{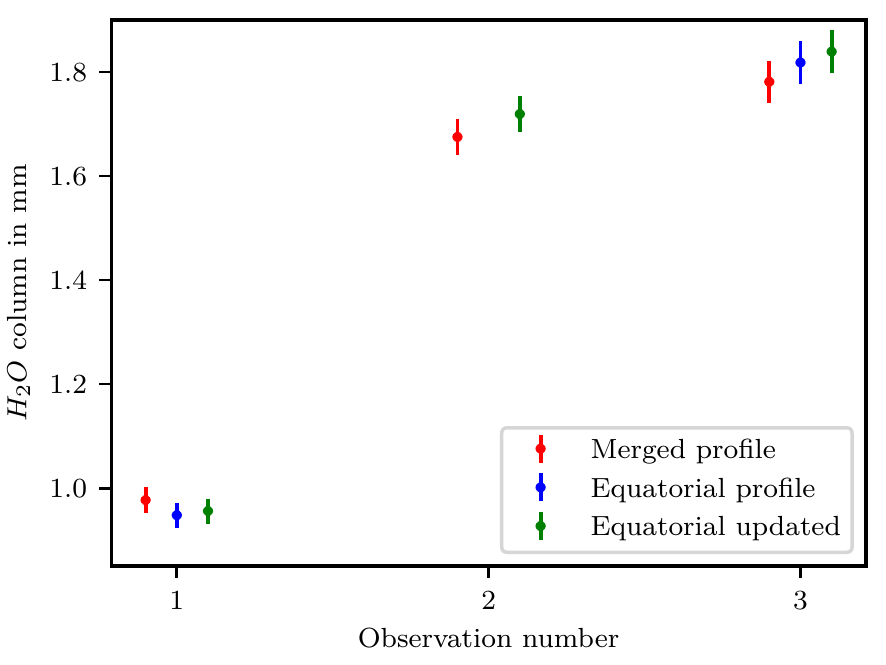}}
  \caption{$H_2O$ column in $mm$ derived from 3 observations of the target star by \molecfit. The red points are obtained using the merged atmospheric profile; the blue points are obtained using the equatorial atmospheric profile. The fit converged to a zero solution for the second observation. Thus, the green points are obtained with the same equatorial profile as the blue points, but the $H_2O$ initial value is modified to that found with the standard star fit.}
  \label{fig:h2o_content}
\end{figure}

We corrected  the target spectra with \molecfit\ as described in Section~\ref{tell_corr}
using the merged profile (EMM + GDAS + equatorial), the equatorial profile,
and the equatorial profile updated with the water volume mixing ratio derived from the standard star. The
\molecfit\ package provides the volume mixing ratio of water in ppmv and the column height of the precipitable water vapour (PWV) in mm.
The column height of PWV is the height of liquid water contained in a vertical column
above the observing site if the water vapour was condensed \citep{marvil_astronomical_2006}.
The calculation to compute the column height of PWV is presented by Equation 9 in \cite{smette_molecfit_2015}. 
We plot in Figure~\ref{fig:h2o_content} the resulting values of the $H_2O$ column obtained with the three atmospheric profiles.
Even though each fit gives a different value, the results are all consistent within their error bars.
We notice that the water column derived by \molecfit\ is increasing with the observation number, contrary to the relative humidity measured at the telescope.

\begin{figure}
  \resizebox{\hsize}{!}{\includegraphics{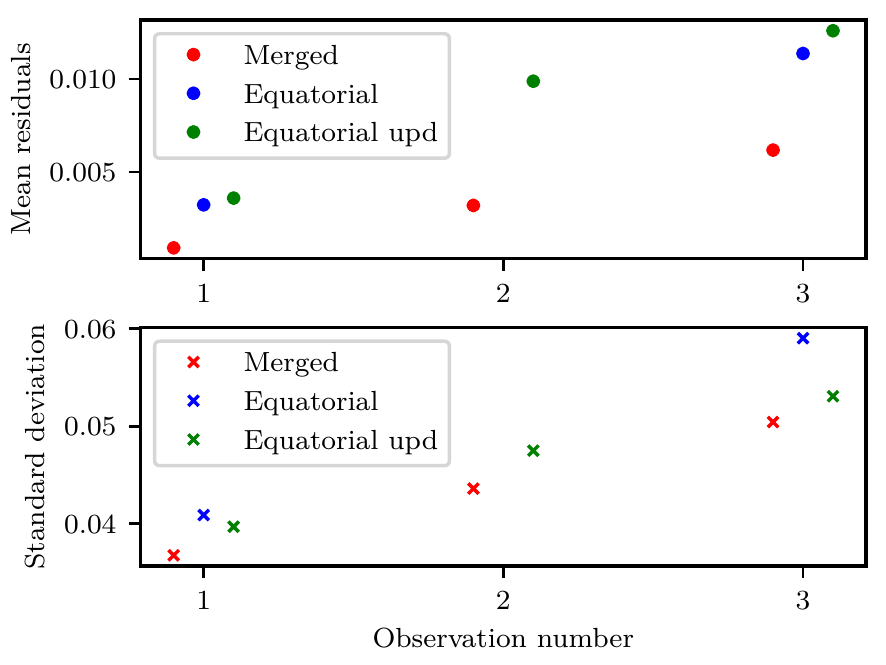}}
  \caption{Mean and standard deviation of the residuals inside the telluric lines. The telluric correction was done with \molecfit\ using three different atmospheric profiles as inputs. The colour code is the same as Figure~\ref{fig:h2o_content}.}
  \label{fig:residual_h2o}
\end{figure}

We measured the residuals between the input spectrum and the \molecfit\ transmission spectrum
inside the telluric lines uncontaminated by stellar lines. Figure~\ref{fig:residual_h2o}
shows the mean and standard deviation of these residuals.
The offset and scatter are increasing with the observation number.
Hence, there is a correlation between the water column derived by \molecfit\ and the accuracy of the telluric correction.

A slight difference in water content within the error bars has an impact on the level of the telluric correction.
Figure~\ref{fig:residual_h2o} also shows that the merged profile always gives smaller mean residuals
than the equatorial profile for the three observations.
We notice that when we updated the starting value of $H_2O$ (green data),
the residuals are closer to those obtained with the merged profile,
which denotes the sensitivity of  \molecfit \ to the starting values.

%
%
%________________________________________________________________

\section{Discussion and conclusions}
\label{conclusion}

We performed telluric correction in the near-infrared on CRIRES data
with three software packages using a synthetic atmospheric transmission.
We also compared these corrections with the standard star method that is commonly used.
The standard star method has several disadvantages, such as the time and airmass differences
between the target and standard star observations, which result in slightly different atmospheric transmission spectra.
Another problem is the loss of telescope time spent observing telluric standard stars.
Using synthetic transmission spectra to model the telluric absorption is a good opportunity
to correct archival data for which no suitable standard star was observed but also to save observing time.

We chose to study the oxygen and water telluric absorption
because these molecules have been used for a long time as wavelength calibrators and have opposite characteristics:
the water absorption varies on short timescales and in strength while the oxygen absorption is quite constant in time and in strength.
For the wavelength range dominated by water absorption,
we found that the synthetic transmission methods deliver a more precise telluric correction.
Inside the water telluric lines, the synthetic transmission methods present systematically lower scatter than the standard star method.
At most, the standard star method has a scatter 1.3 times larger than \molecfit.
We interpret this because water absorption varies quickly in time
and the use of a synthetic transmission of the atmosphere allows us to account for these rapid changes better than the standard star method.

We also found that the constant increase in the scatter of the residuals between observation 1, 2, and 3
is correlated with the increase of the PWV derived by \molecfit.
Deep and narrow water lines might be more difficult to model if the lines are not well sampled.
The correction level obtained with \molecfit\ using any of the atmospheric profile
is still higher than with the standard star method.

For the wavelength range dominated by oxygen absorption the standard star method performs better.
The synthetic transmission methods present a scatter up to 1.2 times higher in the residuals than the standard star method.
Looking more closely inside the telluric lines,
\molecfit\ presents a higher scatter than the standard method but a smaller offset. The
TelFit and TAPAS packages result in a global offset around $1\% $.
For TAPAS, we explain this difference by the fact that part of the $O_2$ absorption
was removed at the time of the continuum normalization.
\cite{gullikson_correcting_2014} also pointed out that TelFit fits with a higher precision
the water lines compared to the oxygen lines.
They attributed it to a probable systematic error in the line strength in the HITRAN database
because two $O_2$ bands were under- and over-fitted by the same mixing ratio at the time.
We cannot test this hypothesis because we have only one $O_2$ band. It might be that the spectroscopic parameters of the $O_2$ molecule require some revisions.

Using additional CRIRES observations, which cover a wider range of airmass and relative humidity,
we compared the three synthetic transmission methods. We found that \molecfit\ and TelFit have
a smaller scatter than TAPAS, especially at low airmass where the scatter obtained with TAPAS is two times higher.
We also showed that an increase in both airmass and relative humidity lead to higher scatter of the residuals for the \molecfit\ and TelFit methods.

On the use of a tailored atmospheric profile for the observatory site,
we showed that the merged atmospheric profile calculated by \molecfit\ always leads to smaller residuals than the equatorial profile.
The merged profile takes into account on-site measurements at Paranal and GDAS atmospheric profiles modelled from a set of meteorological data.
The GDAS profiles are available up to an altitude of 26 km and at higher altitude the merged profile uses the equatorial profile.
The merged profile provides starting values closer to the solution and helps the fit to converge to a solution with a higher precision.

Of the three packages studied, \molecfit\ is the most complete package;
it allows for correction of the telluric absorption across the largest wavelength domain and with the largest number of variables. The
\molecfit\ method is developed by ESO, it is constantly maintained and, at the time of the writing (November 2018), the latest release is dated from September 2018.
The merged atmospheric profile is automatically downloaded and computed by \molecfit.
The large number of variables available to fit the telluric spectra might actually be a disadvantage,
both for the user and the fitting algorithm.
For the user, it might be difficult to pinpoint the right values for each instrument,
while the high number of parameters does not help the convergence of the minimization algorithm.

The TelFit package is an easy to use Python code that can quickly be integrated into an existent code.
We found that calibrating the wavelength solution to the telluric lines was much harder than with \molecfit \ and we were not able to perform the wavelength calibration of the data with the telluric lines.
In addition, as mentioned in \cite{gullikson_correcting_2014}, the scaling by TelFit of the molecular mixing ratios
does not allow for retrieval of the atmospheric parameters, contrary to \molecfit.
The download of the GDAS atmospheric profile is not automated, so we created a script,
available on-line\footnote{\href{https://github.com/soleneulmer/atmospheric_profile}{github.com/soleneulmer/atmospheric\_profile}}, to fill the form and download the profile.

The TAPAS package delivers the atmospheric transmission function,
which has to be fitted to the data to obtained comparable results to the other methods.
The code we wrote to apply this fitting is also available on-line.
It was developed to use with CRIRES data,
but it can be adapted to be used with other spectrographs.
The individual atmospheric transmissions for each molecule are available to download,
information that is essential to fit the abundances of the main absorber.
The TAPAS method does not require any software installation, only a user account.
Disadvantages to TAPAS are that it relies on website availability
and that all the fitting process has to be done after the download of the atmospheric transmission.
The telluric correction using the TAPAS method would certainly benefit
from a fitting procedure being included during the computation
of the atmospheric transmission.

On top of the explored software packages, other similar and available tools could have been used
to perform the telluric correction. The
Tellrem tool uses the same principle as TelFit to compute the atmospheric transmission and to fit the observed spectrum.
However, the line shape fitting can only be done with a Gaussian profile of varying width and the wavelength solution is adjusted with a first degree polynomial.
Overall, Molecfit and TelFit are very similar to this code; this is one of the reasons we chose not to include Tellrem in the comparison.
Also, Tellrem is tailored to be used with X-Shooter spectra, however it could be used to correct spectra from other instruments.
The installation procedure is more complicated than for the two codes because the different components
(radiative transfer code, line database, and atmospheric profiles) all need to be installed separately.

The Planetary Spectrum Generator (PSG; \citealt{villanueva_planetary_2018}) models synthetic planetary spectra
from the ultraviolet to the radio wavelengths.
To do so, PSG uses several radiative transfer codes and spectroscopic databases.
The on-line interface allows for synthesizing the Earth transmittance at the altitude of the Paranal Observatory,
which could be used to model the telluric absorption in the CRIRES observations.
The user can upload parts of their spectrum on the interface
and perform the fitting of the transmission spectrum to the data remotely.

The synthetic atmospheric transmission methods can be limited in case the target star has a lot of intrinsic features,
which complicate the fitting of the telluric lines. The fitting of the synthetic transmission spectra to the observed spectra
should ideally be carried out in regions where there are no blending between the stellar and telluric lines.
The fitting ranges can be small (few tens of angstroms), and stellar lines inside the fitting ranges should be excluded.
However, if no individual telluric line can be identified, the best option might be to keep
the molecular column densities constant and only adjust the wavelength solution for example.
In that case the telluric correction might not lead to the level of correction presented in this work.

The standard star method is a simple and efficient way to correct for the telluric lines,
especially in regions where the water absorption is small. However, the synthetic transmission methods
offer a good alternative to the loss of telescope time spent to observe standard stars
and facilitates obtaining a better precision for the fitting of the water lines.
As a future improvement, it would be important to optimize the fitting algorithms (in particular for \molecfit\ and TelFit)
to explore the parameter space in a more efficient way that is not so dependent on the input values.
It would also be important to compare the different telluric correction methods on a wider wavelength range, as made available by new generation spectrographs.

\subsection*{Acknowledgments}
This work was funded by FEDER - Fundo Europeu de Desenvolvimento Regional funds through the COMPETE 2020 -
Programa Operacional Competitividade e Internacionalização (POCI), and by Portuguese funds through FCT -
Fundação para a Ciência e a Tecnologia in the framework of the project POCI-01-0145-FEDER-032113.
This work was supported by Funda\c{c}\~ao para a Ci\^encia e a Tecnologia (FCT, Portugal)
through national funds and from FEDER through COMPETE2020
by the grants UID/FIS/04434/2013 \& POCI-01-0145-FEDER-007672, and
PTDC/FIS-AST/1526/2014 \& POCI-01-0145-FEDER-016886.
N.C.S. acknowledges support by Funda\c{c}\~ao para a Ci\^encia e a Tecnologia (FCT)
through Investigador FCT contract of reference IF/00169/2012/CP0150/CT0002.
J.J.N. acknowledges support of the FCT fellowship PD/BD/52700/2014.
Plots were done with \texttt{Matplotlib} from \cite{hunter_matplotlib_2007}.
The authors thank Janis Haldeberg for providing the data and Alain Smette for his help with \molecfit.

\bibliographystyle{aa}
\bibliography{atmospheric}

%
%
%________________________________________________________________

\begin{appendix}
  \section{Transmission spectra}
   \label{appendix:trans}

\begin{figure*}
  \sidecaption
  \includegraphics[width=12cm]{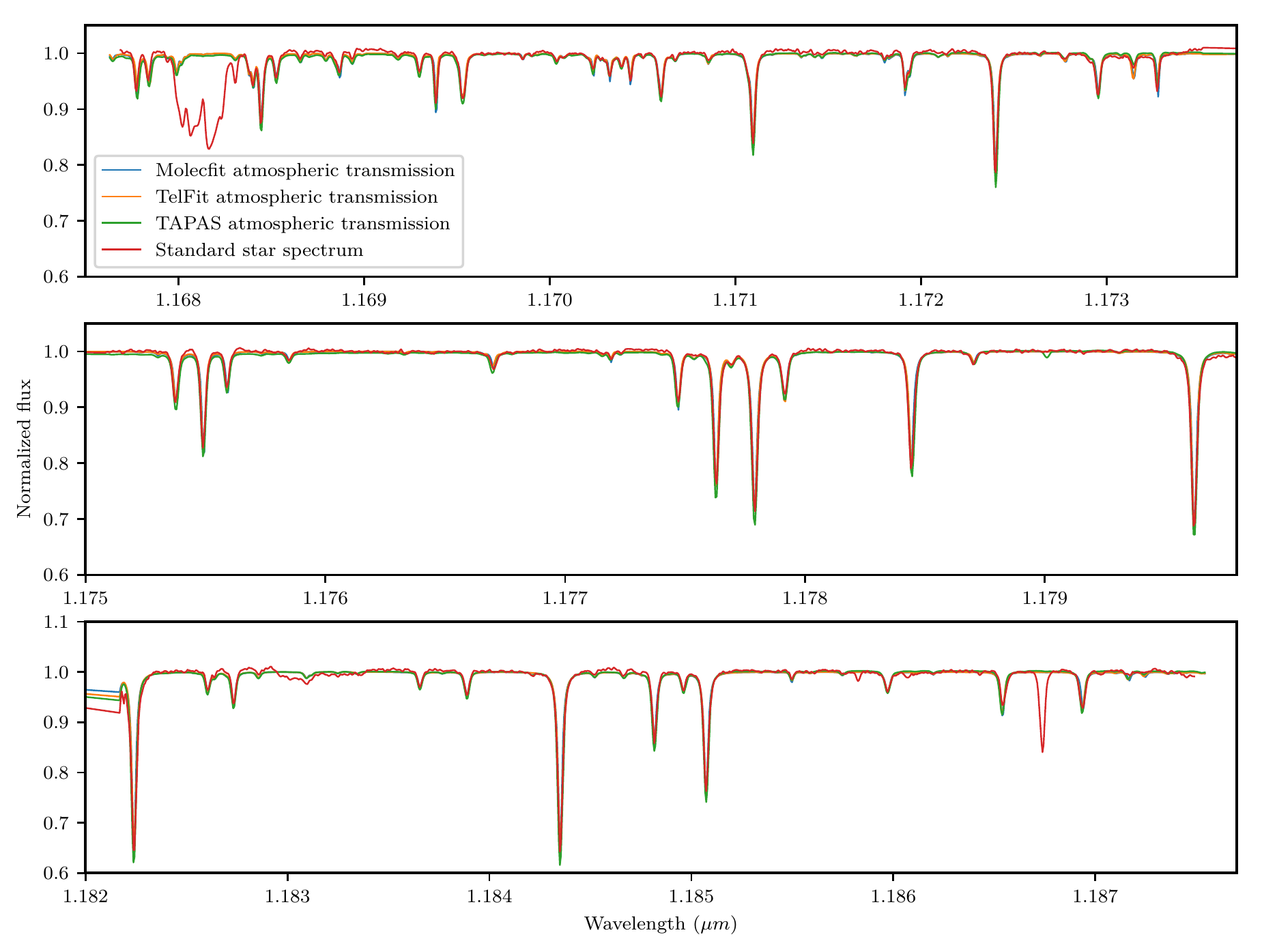}
  \caption{Synthetic transmission spectra and the standard star spectrum dominated by water absorption. The blue curve is computed with \molecfit, the orange curve with TelFit, the green curve with TAPAS, and the red curve is the standard star spectrum.}
  \label{fig:order48}
 \end{figure*}

\begin{figure*}
  \sidecaption
  \includegraphics[width=12cm]{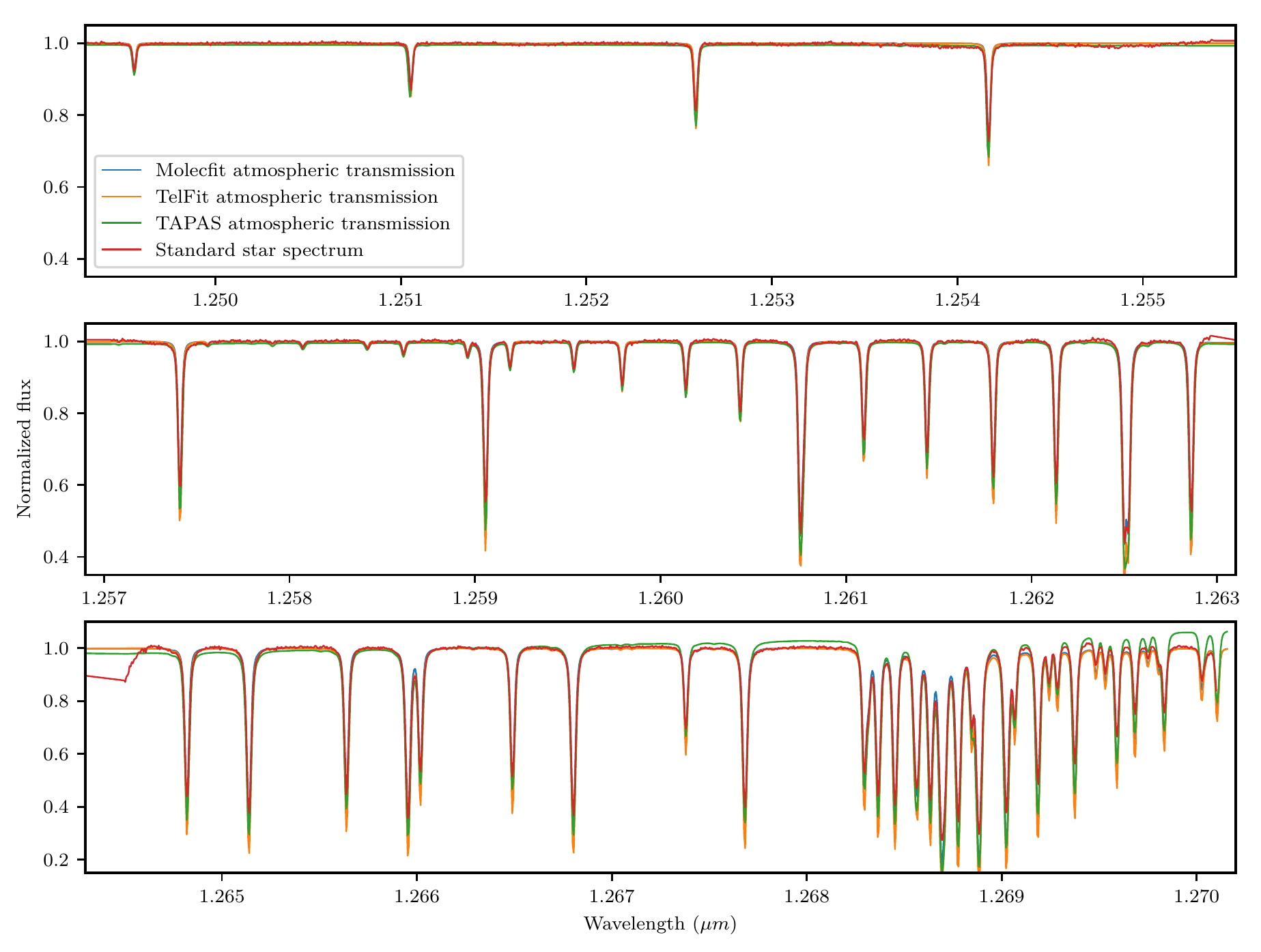}
  \caption{Synthetic transmission spectra and the standard star spectrum dominated by oxygen absorption. The colour code is same as Figure~\ref{fig:order48}}
  \label{fig:order45}
 \end{figure*}

   This appendix presents the three synthetic transmission spectra obtained
   with \molecfit, TelFit, and TAPAS as well as the standard star spectrum.
   Figure~\ref{fig:order48} shows the wavelength range from
   \SI{1.166}{\micro\metre} to \SI{1.188}{\micro\metre}, where water absorption dominates.
   The standard star spectrum has lines remaining that are not matched by the synthetic transmission spectra.
   The remaining lines at \SI{1.168}{\micro\metre} are probably due to a detector problem since previous spectra
   of the same star do not show these lines. The line at \SI{1.1867}{\micro\metre} resembles a narrow telluric line,
   but it is not reproduced by any of the synthetic atmospheric transmission spectra.
   Figure~\ref{fig:order45} shows the second wavelength range from \SI{1.247}{\micro\metre} to \SI{1.272}{\micro\metre}, where oxygen absorption dominates.
   In this figure, we do not notice remaining lines in the standard star spectrum. The major difference is the continuum level of each synthetic transmission spectra.

 \section{Telluric corrections}
 \label{appendix:trans2}
\begin{figure*}
  \sidecaption
  \includegraphics[width=12cm]{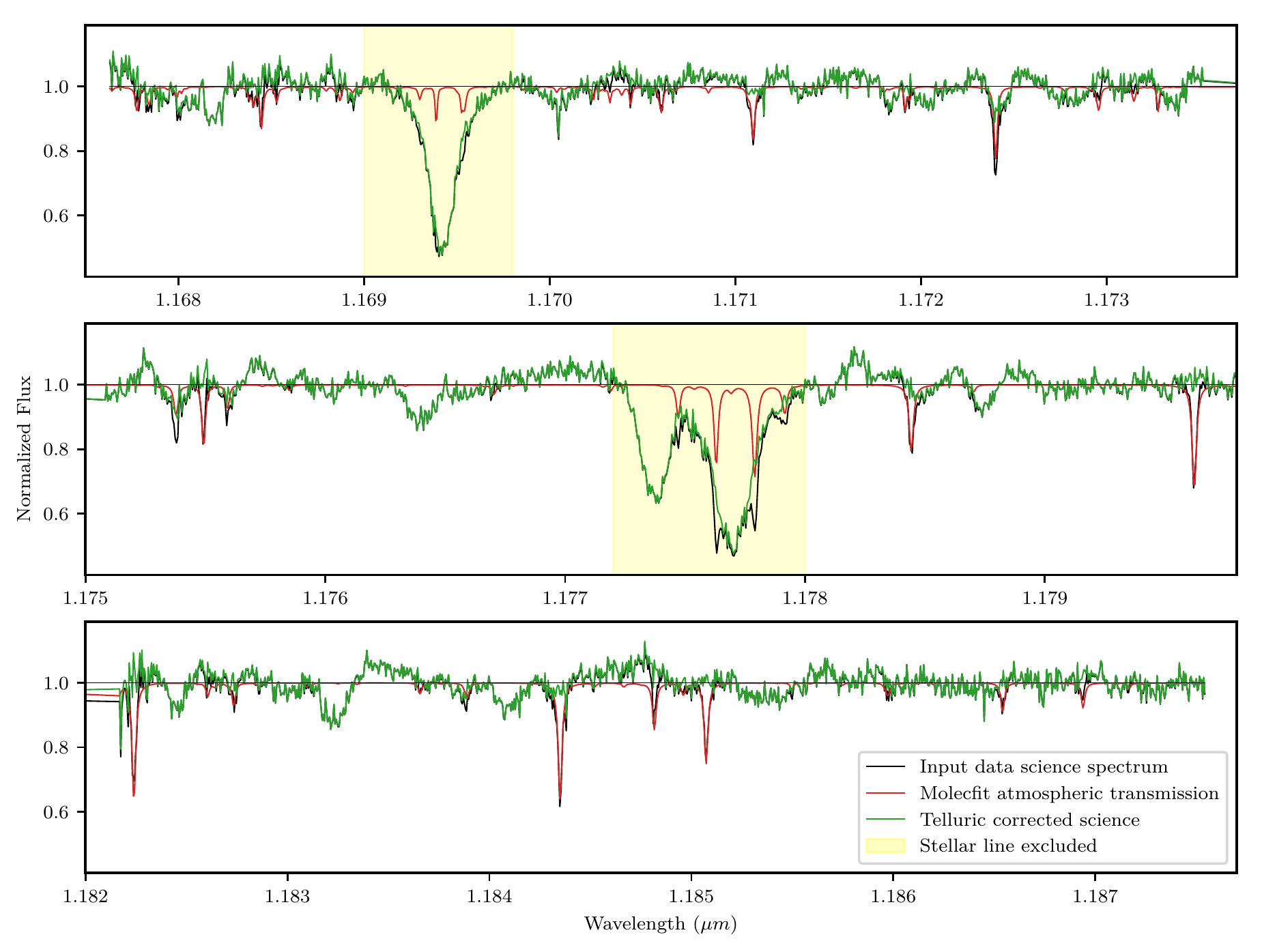}
  \caption{Example of a telluric-corrected spectrum with \molecfit. The input spectrum is indicated in black, the atmospheric transmission spectrum is indicated in red, and the telluric-corrected spectrum is indicated in green.}
  \label{fig:molec0}
\end{figure*}

\begin{figure*}
  \sidecaption
  \includegraphics[width=12cm]{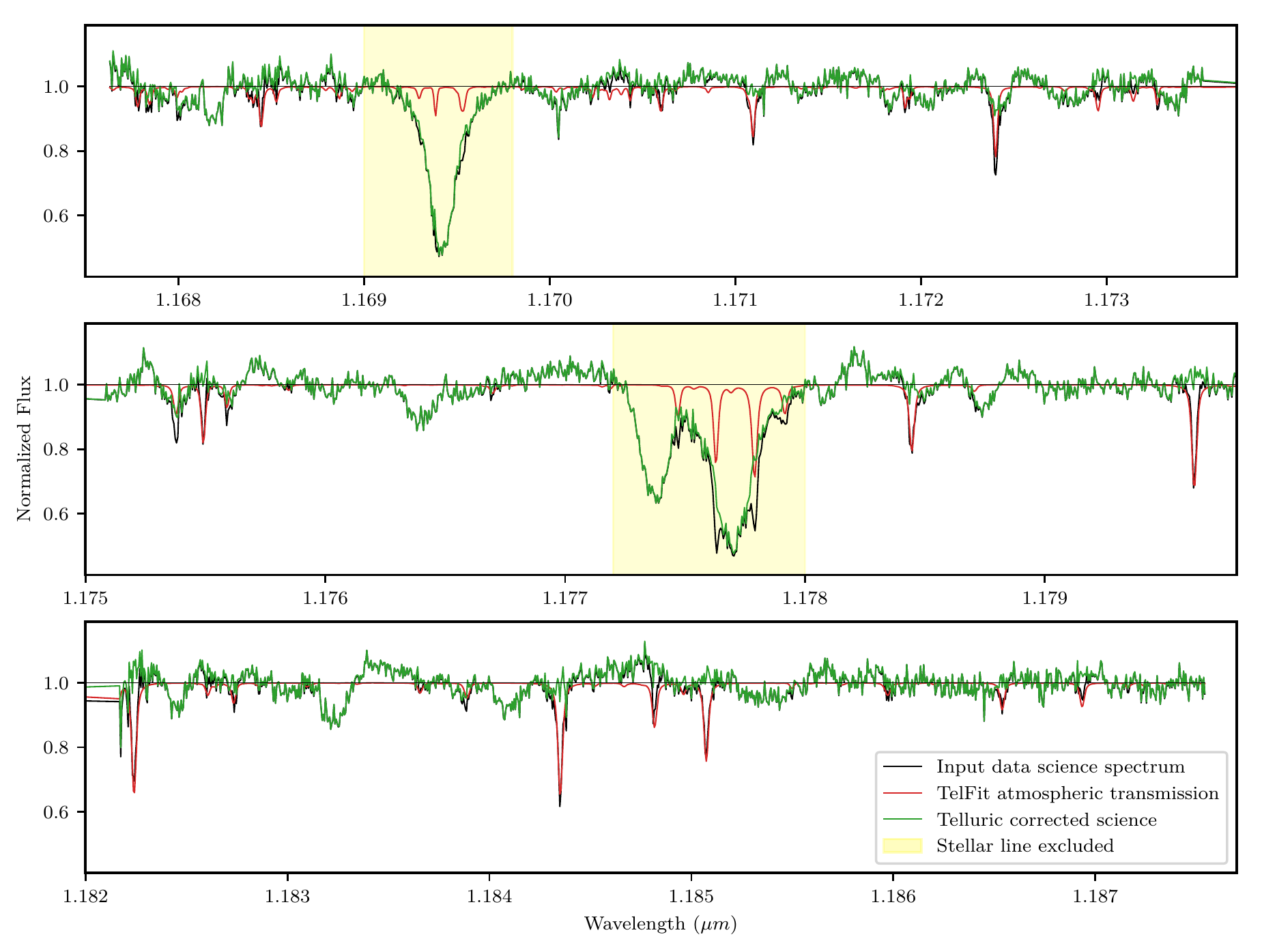}
  \caption{Example of a telluric-corrected spectrum with TelFit. The colour code is same as Figure~\ref{fig:molec0}}
  \label{fig:telfit0}
\end{figure*}

\begin{figure*}
  \sidecaption
  \includegraphics[width=12cm]{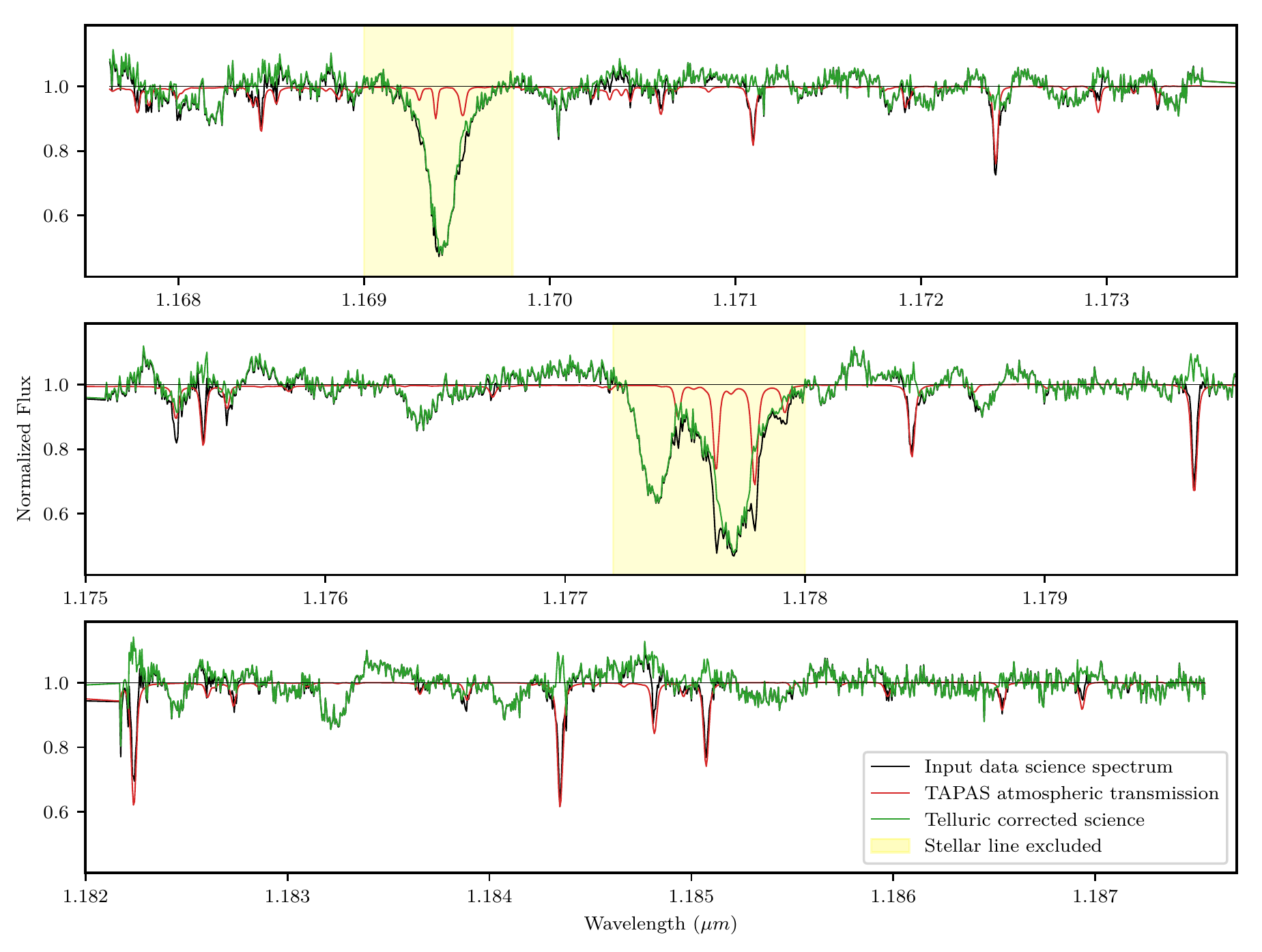}
  \caption{Example of a telluric-corrected spectrum with TAPAS. The colour code is same as Figure~\ref{fig:molec0}}
  \label{fig:tapas0}
\end{figure*}

\begin{figure*}
  \sidecaption
  \includegraphics[width=12cm]{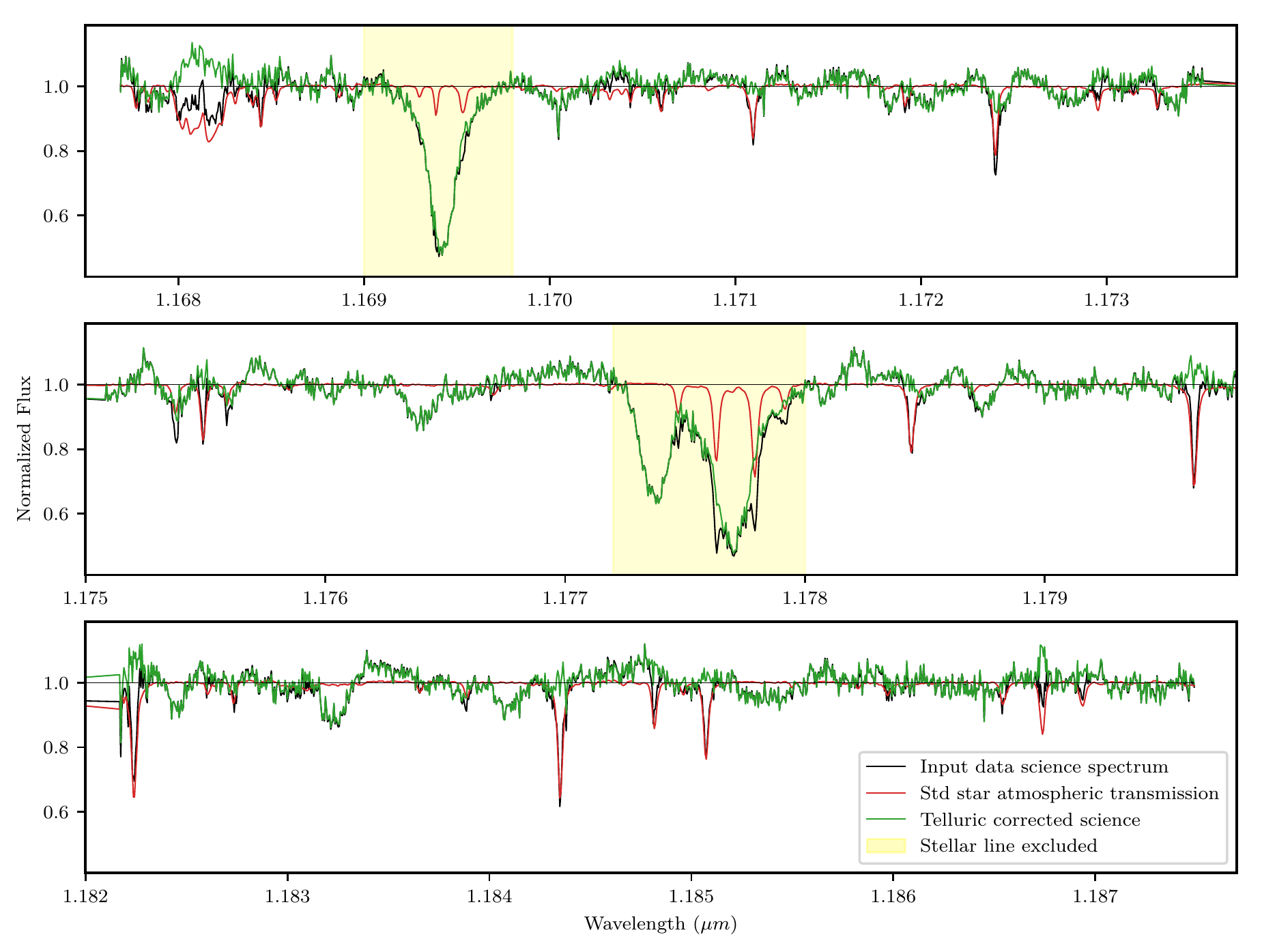}
  \caption{Example of a telluric-corrected spectrum with the standard star method. The colour code is same as Figure~\ref{fig:molec0}}
  \label{fig:std0}
\end{figure*}

\begin{figure*}
  \sidecaption
  \includegraphics[width=12cm]{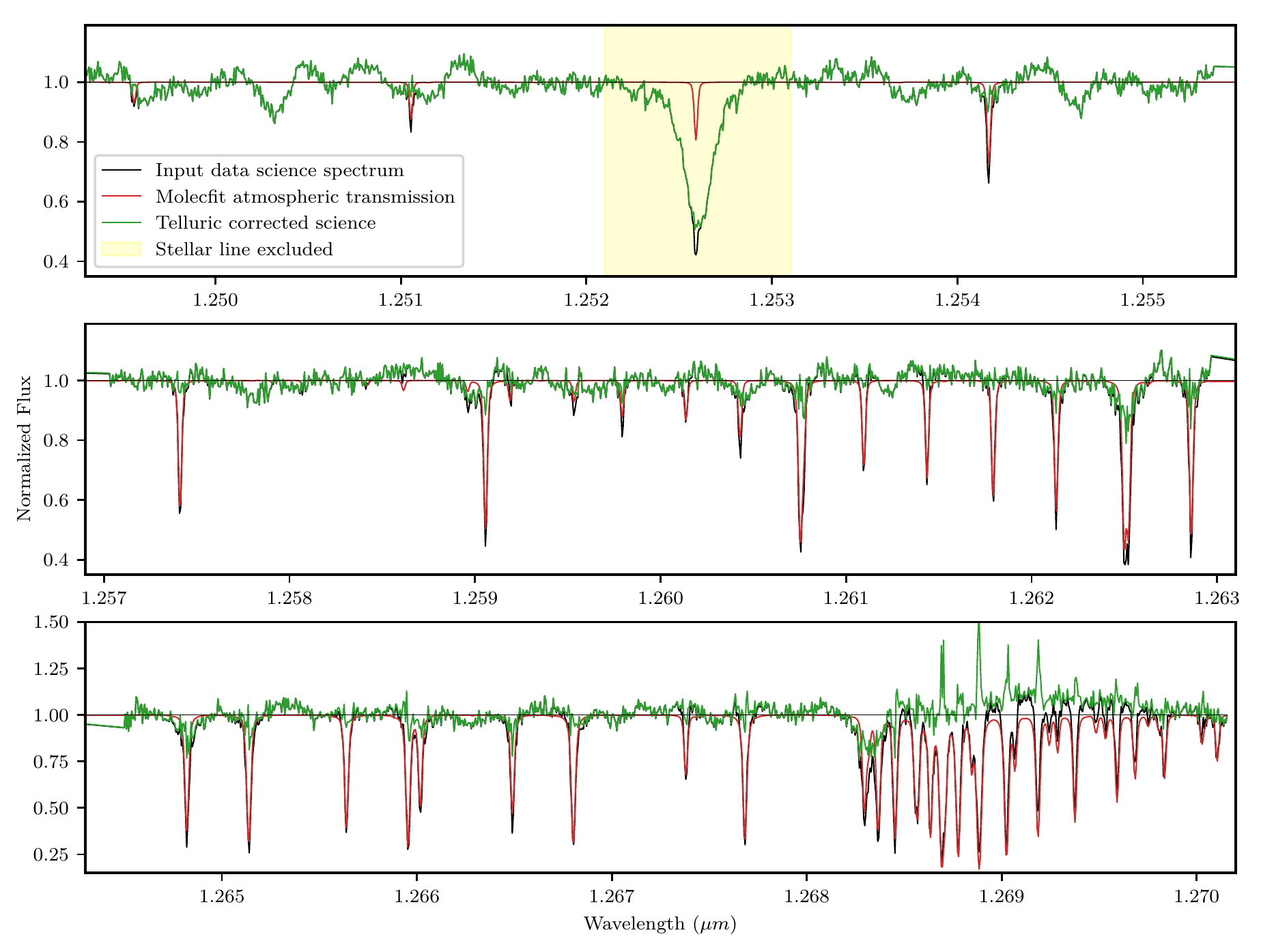}
  \caption{Example of a telluric-corrected spectrum with \molecfit. The colour code is same as Figure~\ref{fig:molec0}}
  \label{fig:molec1}
\end{figure*}

\begin{figure*}
  \sidecaption
  \includegraphics[width=12cm]{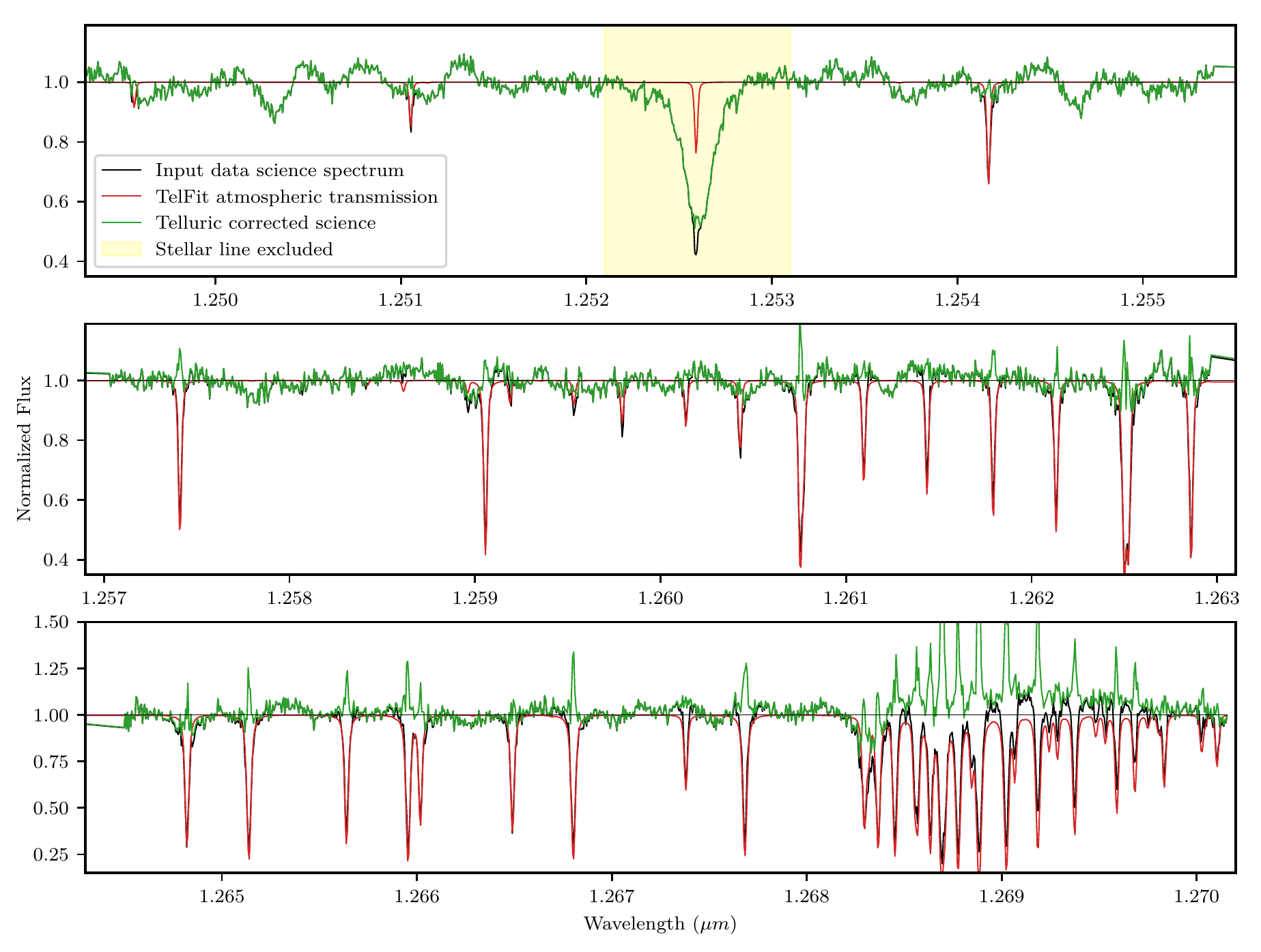}
  \caption{Example of a telluric-corrected spectrum with TelFit. The colour code is same as Figure~\ref{fig:molec0}}
  \label{fig:telfit1}
\end{figure*}

\begin{figure*}
  \sidecaption
  \includegraphics[width=12cm]{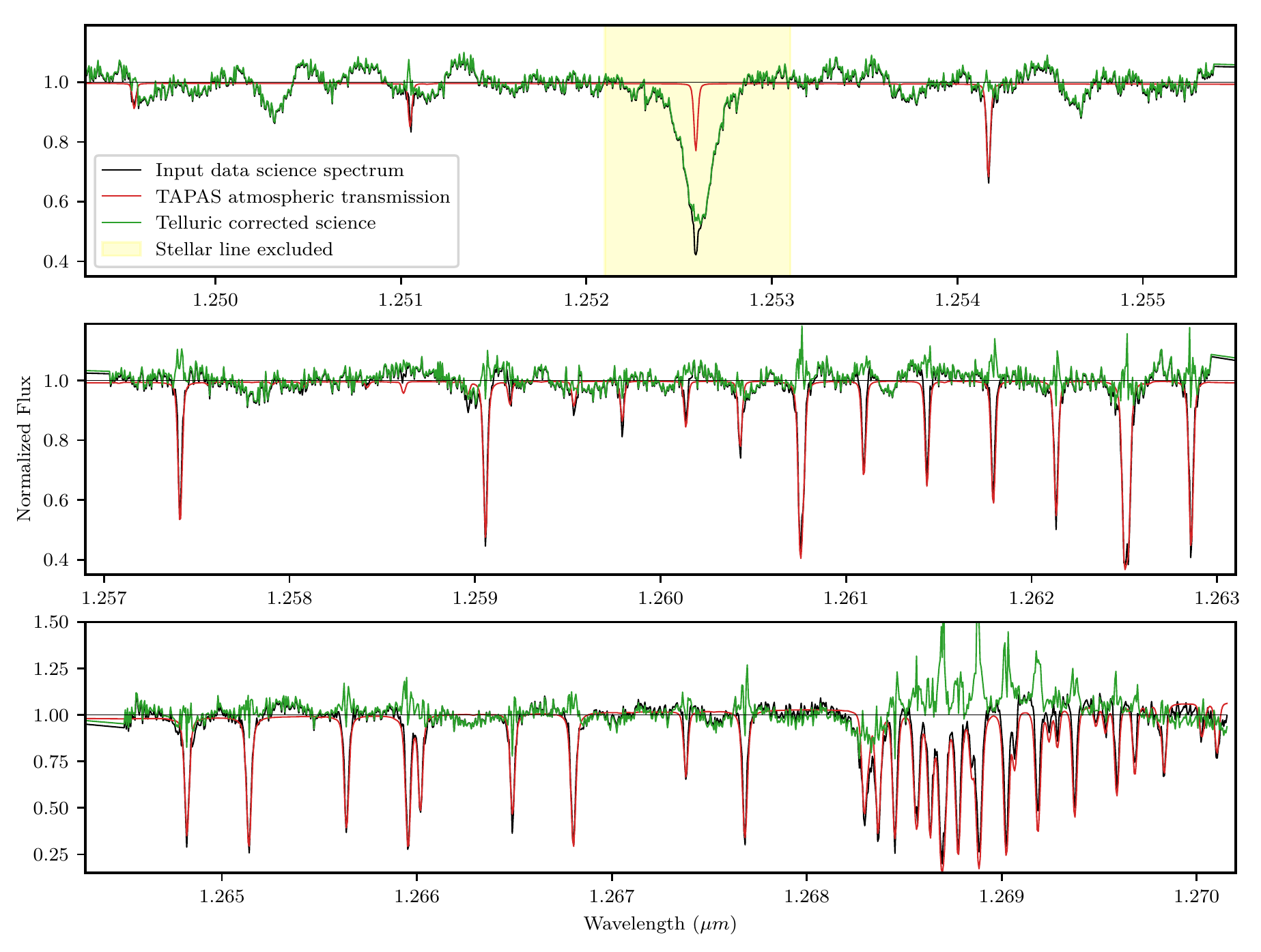}
  \caption{Example of a telluric-corrected spectrum with TAPAS. The colour code is same as Figure~\ref{fig:molec0}}
  \label{fig:tapas1}
\end{figure*}

\begin{figure*}
  \sidecaption
  \includegraphics[width=12cm]{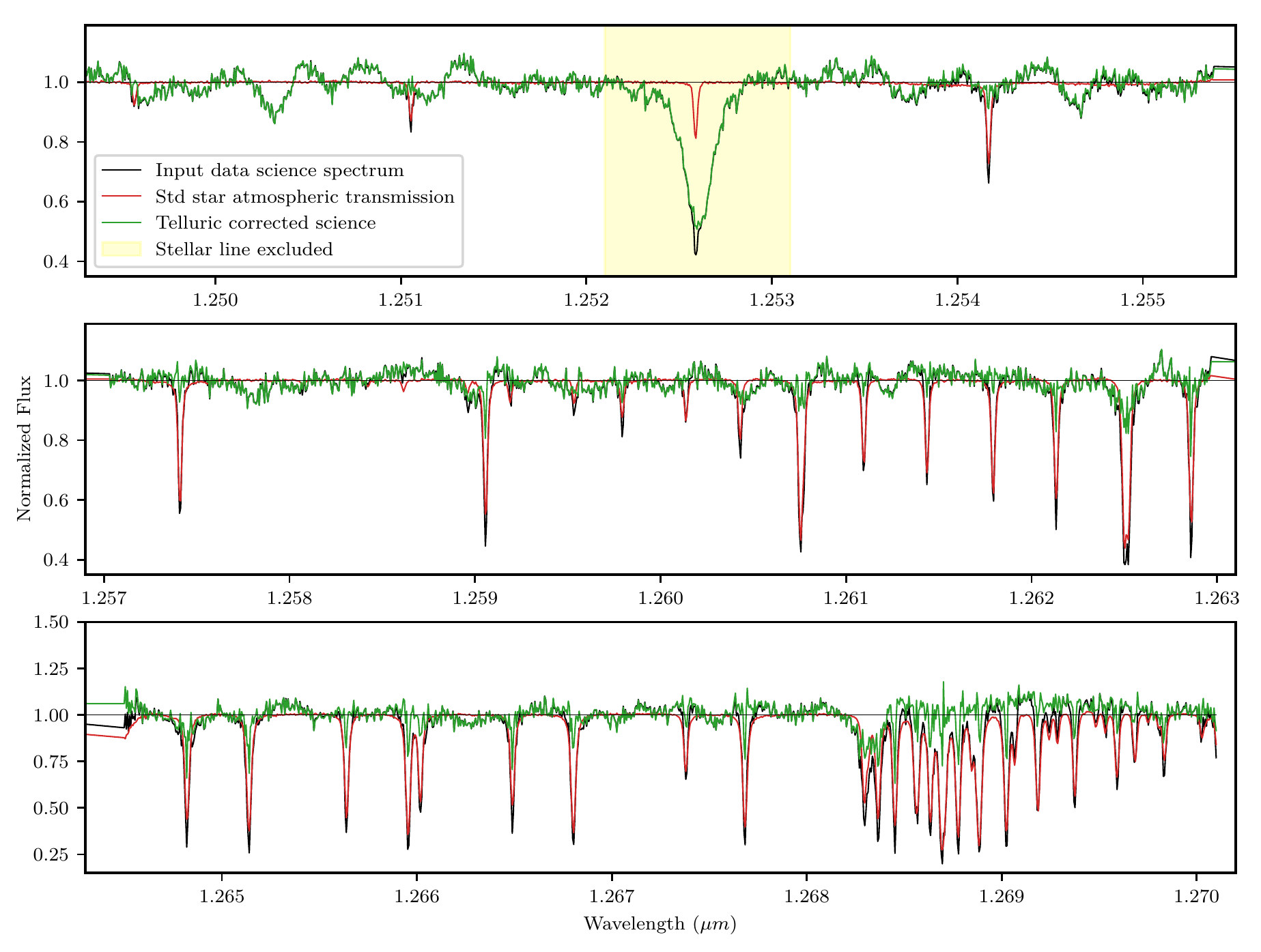}
  \caption{Example of a telluric-corrected spectrum with the standard star method. The colour code is same as Figure~\ref{fig:molec0}}
  \label{fig:std1}
\end{figure*}

 We present the plots of the telluric correction in both wavelength ranges. The telluric correction was performed with \molecfit\ in Figures~\ref{fig:molec0} and \ref{fig:molec1}, TelFit in Figures~\ref{fig:telfit0} and \ref{fig:telfit1}, TAPAS in Figures~\ref{fig:tapas0} and \ref{fig:tapas1}, and the standard star method in Figures~\ref{fig:std0} and \ref{fig:std1}. 
 
\end{appendix}

\vfill
\eject
\end{document}